\newcommand{\du}{\,cm$^{-3}$\,}
\newcommand{\pu}{\,dyn~cm$^{-2}$\,}
\newcommand{\lu}{\,erg s$^{-1}$\,}

\documentclass[12pt,preprint2]{aastex}
\usepackage{natbib}

\shortauthors{Maggio et al.}
\shorttitle{X-ray spectroscopy of AD Leo with {\it Chandra}}

\received{2004 March 3}
\accepted{2004 May 26}
\begin{document}

\title{X-ray spectroscopy of the unsteady quiescent corona of AD Leo
with {\it Chandra}}

\author{A. Maggio\altaffilmark{1}, J.J. Drake\altaffilmark{2}, 
V. Kashyap\altaffilmark{2}, F.R. Harnden, Jr.\altaffilmark{2},\\
G. Micela\altaffilmark{1}, G.~Peres\altaffilmark{3},  
S. Sciortino\altaffilmark{1}}

\altaffiltext{1}{INAF -- Osservatorio Astronomico di Palermo Giuseppe S. Vaiana,
Piazza del Parlamento 1, I--90134 Palermo, Italy;
maggio@astropa.unipa.it, micela@astropa.unipa.it, sciorti@astropa.unipa.it.}
\altaffiltext{2}{Smithsonian Astrophysical Observatory,
60 Garden Str., 02138 Cambridge MA, USA; jdrake@cfa.harvard.edu,
vkashyap@cfa.harvard.edu, frh@cfa.harvard.edu.}
\altaffiltext{3}{Dipartimento di Scienze Fisiche ed Astronomiche -- Sezione di
Astronomia -- Universit\`a di Palermo, Piazza del Parlamento 1, I--90134
Palermo, Italy; peres@astropa.unipa.it.}

\begin{abstract}

We present the analysis and interpretation of an observation
of the flare star AD Leo (dM3e) with the Low-Energy Transmission
Grating of {\em Chandra}. The high resolution X-ray spectrum --
dominated by emission lines from O\,VII--VIII, Ne\,IX--X, and
Fe\,XVII--Fe\,XIX -- allowed us to infer the plasma emission measure
distribution (EMD) vs.\ temperature, and the abundances of individual
elements in the corona of this magnetically-active star, during 
a typical state characterized by significant variability but no evident
flaring event. We have also measured plasma densities at various
temperatures using spectroscopic diagnostics provided by
He-like triplets and \ion{Fe}{21} lines. We show that the present EMD
is similar, in terms of overall shape and temperature of the peak, to those previously obtained from EUVE spectra 
during quiescent and flaring states confirming the long-term stability of the
corona of AD~Leo. At variance with the case of other active stars,
the EMD of AD~Leo is characterized by a significantly shallower slope,
compatible with that predicted by static models of isobaric loops with
constant cross-section and uniform heating. We discuss such coronal
modeling to infer the average properties of the corona in terms of loop
populations, including estimates of the surface filling factor derived
by comparison between the model and the observed EMD. We also show that
the EMD is compatible with the model of a corona continuously heated by
flares, which predicts an EMD slope slightly steeper than observed,
but that can be accommodated by observational uncertainties. 
The coronal composition is such that the element abundances, relative to
solar values, tend to increase with the First Ionization Potential, with
few exceptions. The line to continuum ratios suggest a
nearly solar metallicity, a result difficult to reconcile with previous
determinations based on global fitting of X-ray spectra.

\end{abstract}

\keywords{stars: individual (AD Leo) --- stars: activity --- stars: coronae
--- stars: late-type --- X-rays: stars}

\section{Introduction}
\label{sec:intro}

One of the main challenges of stellar astrophysics today is 
to infer the characteristics of the magnetic structures
that confine hot plasma in stellar coronae 
and the nature of coronal heating. 
The topology, surface coverage and strength of the coronal
magnetic fields are crucial factors in determining the amount of mass
and angular momentum losses via magnetically-coupled stellar winds; 
such determinations may further our understanding of stellar rotational
history, the effects of rotation on stellar evolution and
the role normal stars play in the metal enrichment of the interstellar medium 
and in the chemical evolution of our Galaxy.
Determining the plasma temperature 
and density distributions in stellar coronae, as well as the manner in which
these properties change with the X-ray activity level, are key to this understanding.

In this context, active dMe stars are very appealing objects of study because
they likely represent the most numerous class of galactic soft X-ray sources,
and they allow us to study stellar coronae possibly quite 
different from the Sun's. 
In fact, almost all such stars emit X rays in the
so-called saturation limit \citep{pmm+03}, 
having a ratio of X-ray to bolometric
luminosity $L_{\rm x}/L_{\rm bol} \approx 10^{-3}$ (compared to 
$L_{\rm x}/L_{\rm bol} \approx 10^{-6}$ for the Sun), Moreover, the
X-ray emission of these stars is highly variable on short time scales
(from hundreds of seconds to days) due to the occurrence of many
flaring events of different intensities, which might represent
one of the main mechanisms for plasma heating \citep{agdk00,kdga02,g+03},
Furthermore, many of these
stars are relatively fast rotators compared to the Sun,
their convective regions are much deeper and hence their magnetic
dynamo activity likely operates on different spatial and temporal scales.

Our present understanding of the corona of AD Leo and a few other dMe
stars takes advantage of four different methodologies applied to 
existing X-ray and EUV data: global spectral fitting of low- or
medium-resolution X-ray spectra, analysis of flare decays, time
variability studies and plasma emission measure analysis from
high-resolution emission line spectra.

X-ray spectra obtained with {\it ROSAT} 
and {\it BeppoSAX} were previously employed
to infer the properties of the coronal loop-like structures
\citep{grk+96,smf99} by comparison with theoretical models of
the magnetically-confined X-ray emitting plasma \citep*{rtv78,spv+81}, 
in analogy with the case of the solar corona.  Such modeling
provides estimates of basic parameters like the scale size of the
coronal loops, the plasma pressure and the fraction of the stellar surface 
covered by the emitting regions.  In practice, different combinations of
these quantities are usually
allowed by the data, due to a fundamental degeneracy of the solutions
of the models when the structures are smaller than the pressure scale
height \citep{mp96}. 

More robust constraints on the sizes of {\it individual} coronal loops
have been derived from a detailed analysis of the decay
phase of prominent flaring events, 
often displayed by active red dwarfs \citep{rm98,frm+00,gar+04};
extrapolation of the results of this analysis to the entire corona
relies on the assumptions that the flaring loops are ``typical''
coronal structures and that the maximum temperature of the plasma in
the same structures during quiescent phases can be guessed by fitting
low- or medium-resolution X-ray spectra (taken before or after the flares) 
with simple models that contain a few isothermal components. In fact, the
knowledge of the loop size (derived from the flare decay) and of the
plasma maximum temperature and total emission measure (derived from the
spectral fitting) permits computation of the other relevant parameters
(e.g., plasma pressure, surface filling factor) as explained in
\citet{smf99}.

Complementary information on the nature of stellar coronae
is provided by variability analysis of X-ray or EUV light
curves. In particular, the coronal emission of red dwarf stars has been studied
by \citet*{asg87} using {\it Einstein} data, by \citet*{cps88} with
EXOSAT, by \citet{grk+96} with
ROSAT, and more recently by \citet{kdga02} with EUVE observations.
The main result of these studies is that significant
variability is continuously observed, and likely associated with
flaring events spanning a large range of possible intensities (i.e.,
released energy); these events might explain the steady presence of
hot plasma in the coronae of active stars, by
contributing a significant fraction of the total coronal heating.

In this paper we analyze and interpret an observation of the dMe flare star 
AD~Leo obtained with the Low-Energy Transmission Grating Spectrometer 
\citep[LETGS,][]{b+01} on board the {\it Chandra} X-ray Observatory 
\citep{wbc+02}.
AD~Leo has been observed by virtually all previous space-borne X-ray 
observatories, but the observation discussed here
provided the first high-resolution X-ray line spectra
from which several diagnostics on the temperature and density structure
of the coronal plasma and its chemical abundances can be
self-consistently derived. The new information provided by this
{\it Chandra} observation has allowed us to test and improve our understanding
of the corona of AD~Leo, and to delineate a comprehensive reference scenario
for future analyses of the coronal emission of other active M-type dwarfs.

In Sect.~\ref{sec:targ} we introduce our target and its previous observations, 
and in Sect.~\ref{sec:data} our data reduction approach.
Sect.~\ref{sec:a+r} is
devoted to the results of our analysis, with most of the technical
details relegated to the Appendix. We discuss these results in
Sect.~\ref{sec:discuss} and draw our conclusions in Sect.~\ref{sec:concl}.

\begin{deluxetable}{ccccccc}
% \tablewidth{0pt}
\tablecaption{Target characteristics \label{tab1}}
\tablehead{
\colhead{Target} & \colhead{Parallax} & \colhead{Spectral} & 
\colhead{} & \colhead{Metallicity} & \colhead{Mass} & \colhead{Radius}\\
\colhead{Name} & \colhead{(mas)} & \colhead{Type} & 
\colhead{$M_{\rm bol}$} & \colhead{[Z/H]} & 
\colhead{$M_{\sun}$} & \colhead{(cm)}
}
\startdata
AD Leo (GJ 388) & $213 \pm 4$ & M3.5V & 8.85 & $-0.75 \pm 0.25$ & 0.40 & $2.6
\times 10^{10}$ \\
\enddata
\tablerefs{Parallax by \citet{j52};
metallicity by \citet{jlah96}; other parameters by \citet{fmr00}}
\end{deluxetable}

\section{Previous observations of AD~Leo}
\label{sec:targ}

AD Leo is one of the brightest X-ray coronal sources among the {\em single} 
dMe stars currently known. It is a nearby star (D = 4.7 pc), with an intervening
interstellar hydrogen column density $N_{\rm H} = (3 \pm 1) \times
10^{18}$\,cm$^{-2}$, determined from previous observations
in the EUV band \citep{cfh+97}. 
Other characteristics of this object are reported in Table~\ref{tab1},
and further discussion of its fundamental parameters (mass $M = 0.4
M_{\sun}$, radius $R \sim 0.37 R_{\sun}$, bolometric luminosity $L_{\rm
bol} = 8.7 \times 10^{31}$\lu) can be found in \citet{fmr00}.

Several variability analyses of the coronal emission of AD~Leo 
were performed in the past (see Sect.~\ref{sec:intro} for a brief review).
Most recently,
\citep*{fmr00} presented an extensive study of X-ray flares on AD~Leo,
whose results will be discussed in Sect.~\ref{sec:phys}, while
\citet{sm02} and \citet{g+03} have analyzed several EUVE observations.

The quiescent X-ray luminosity of AD~Leo was found by \citet{fmr00} to be
quite stable at $L_{\rm x} = 3$--$5 \times 10^{28}$\,erg s$^{-1}$
(0.5--4.5\,keV band),
in earlier X-ray observations with {\it Einstein}, {\it ROSAT},
and {\it ASCA} that covered a period of 17\,yr 
(1980--1997). On shorter time scales, however, AD~Leo is strongly
variable: from data taken with the Deep Survey monitor on EUVE, 
\citet{agdk00} have estimated a frequency of 7 flares per day with
total released energy $E \gtrsim 5 \times 10^{31}$\,erg, and 1 flare per
day with $E \gtrsim 5 \times 10^{32}$\,erg, all above a quiescent EUV
luminosity of $\sim 9 \times 10^{28}$\, erg s$^{-1}$.

An attempt has been made by \citet{g+03} to predict the time-averaged
plasma emission measure distribution (EMD) vs.\ temperature in the
corona of AD Leo, based on the assumption that the corona is heated by
a population of flares with a power-law energy distribution: the
results of this exercise appear encouraging but have not yet met the
test of an EMD derived from a high signal-to-noise, 
high-resolution emission line spectra.

\begin{deluxetable}{ccccc}
\tablecaption{Observation characteristics \label{tab2}}
\tablehead{
\colhead{Chandra} & \colhead{} & \colhead{Start} & \colhead{Stop} &
\colhead{Livetime} \\
\colhead{configuration} & \colhead{ObsId} & \colhead{Time} & \colhead{Time} & 
\colhead{(s)}}
\startdata
LETG + HRC-S &  24 & 2000 Jan 22 20:22:42 & 2000 Jan 22 23:37:30 &  9147 \\
LETG + HRC-S & 975 & 2000 Oct 24 15:06:16 & 2000 Oct 25 04:48:37 & 48105 \\
\enddata
\end{deluxetable}

Prior to the X-ray analysis presented here, spectra with
resolution adequate for a detailed emission measure analysis had been
provided by a number of EUVE observations discussed by \citet{cfh+97}
and by \citet{sm02}. These analyses, based essentially on iron lines
alone, suggested a complex coronal structure with several possible
classes of coronal loops present at any time. Available line
ratio diagnostics indicated the presence of very high density plasma
($N_{\rm e} \sim 5$--$8 \times 10^{12}$\du) at temperatures of 10 MK, 
even when evident flaring episodes were removed from the data stream.

\section{{\it Chandra} Observations and Reduction}
\label{sec:data}

AD~Leo was observed twice as part of the {\it Chandra} GTO program.
A first, relatively short observation was followed by a second
much longer observation made nine months after the first, as
summarized in 
Table~\ref{tab2}. Preliminary results of the analysis of each
observation have been presented by \citet{mdh+01,mdk+02}. 
Due to the low signal-to-noise (S/N) ratio of the data in
the first observation, the present work is based primarily on the 2000
October observation. This observation was recently considered by
\citet{brm+03}, who performed a partial analysis of two separate data
segments with the aim of determining the physical
conditions of the corona of AD~Leo in different states. 
Their analysis suggests primarily that only
the high-temperature tail of the EMD was enhanced during the first segment,
but the lack of statistical uncertainties on the EMD values prevents to assess 
the significance of such enhancement. 
Since some variation in the amount of hot ($T > 10^7$\,k) plasma
is an inherent and expected characteristic of such an active star,
we feel that a more thorough analysis of the entire second observation is 
not only justified but also demanded to fully exploit the highest 
S/N ratio achievable with the available data.

We have re-processed the data with the {\it Chandra} Interactive
Analysis of Observations (CIAO V2.2) software and relevant science
threads\footnote{\url{http://asc.harvard.edu/ciao/threads}}.
In order to optimize the S/N ratio of the target spectrum,
this software includes position-dependent PHA filtering of the data
(which reduces the background level by a factor $\sim 4$)
and extraction of the (raw) source spectrum from an optimized bow-shaped 
region, ensuring that a nearly uniform fraction (90\% to 94\%) of source photons, 
is collected at all wavelengths. 
The background spectrum is obtained as the sum of two spectra
extracted from symmetrical regions above and below the source
region, with each having a uniform area five times that of
the source area, at all wavelengths.

We computed our net spectrum
by subtracting a background spectrum -- smoothed with a boxcar filter four
pixels wide and scaled to the source region extraction width -- from the source spectrum, and then co-adding the positive
and negative orders. 

% \clearpage
\begin{figure}[!ht]
\plotone{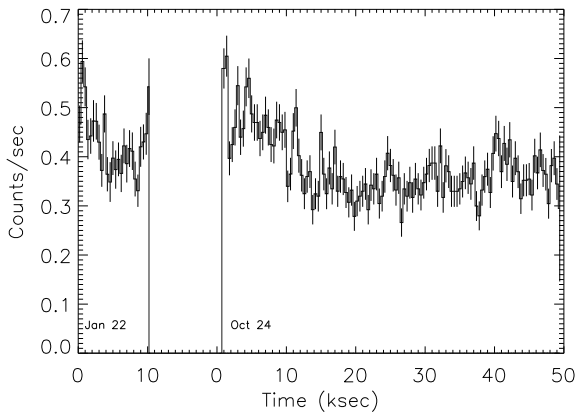}
\caption{
AD Leo X-ray light curve (net count rate)
extracted from the 0-th order source region.
Although no major flare occurred significant
variability is present at the 99\% statistical confidence level.
Bin size is 400 sec, except for the $\sim$9-month gap (Jan 22 to Oct 24)
between the two observations.
\label{fig:lc}}
\end{figure}

\section{Analysis and Results}
\label{sec:a+r}

\subsection{X-ray light curve}
\label{sec:lc}

Figure~\ref{fig:lc} shows the X-ray light curve obtained using the
0th-order source photon arrival times, after subtraction of background 
collected in an annulus adjacent to the 0th-order source extraction region. 
No strong isolated flare is visible during the observation, but some low-level
variability is clearly present: as quantified by a simple $\chi^2$ test,
the light curve deviates from the observed mean count rate (0.38 cnt s$^{-1}$)
at the 99\% confidence level.
The light curve
shows a slowly declining X-ray emission level during the first 12\,ksec 
of the Oct 2000 observation, but we are unable to distinguish whether 
this is the final decay phase of a flare or some other modulation effect; 
in fact, the likelihood of a flare is very high, 
given their observed rate of occurrence (Sect.~\ref{sec:targ}), 
but a rotational modulation effect cannot be completely excluded because 
the duration of the October observation is a 
sizeable fraction ($\sim 20$\%) of the photometric rotational period 
($P_{\rm rot} = 2.7$\,d, Spiesman \& Hawley 1986).
In any case, such variability is typical of the ``unsteady
corona" of AD~Leo, and given the absence of any large
variation of the emission level, 
we have chosen to analyze the spectral data accumulated over the 
entire length of both observations in order to maximize S/N ratio.
A more detailed study of the variability of AD Leo and other active
stars observed by {\it Chandra} will be presented in a forthcoming
paper (Argiroffi et al. 2003, in preparation).

\subsection{Spectral analysis}
\label{sec:sp}

The spectral analysis was performed using the software Package 
for INTeractive Analysis of Line Emission 
\citep[PINTofALE,][]{kd00} 
with an iterative procedure including the following steps:
\begin{enumerate}
\item 
Lines were identified on the basis of the {\sc chianti} V4.02 \citep{dly+01}
database and a model spectrum synthesized from a trial
EMD and trial solar element abundances \citep*{gns+92},
with line emissivities computed by adopting the ionization balance of 
\citet{mmcv98}. Identifications were subsequently verified with the 
final EMD and estimated abundances.
\item
Line profiles were fit with an analytical approximation to the instrumental line spread
function (Sect.~\ref{sec:id}) plus a piece-wise constant base level
(required to model the source continuum,
which was found to be nearly flat.
After the first iteration of the analysis, the continuum 
was adjusted where required to match that predicted by the
computed EMD, and the line fitting was
repeated if necessary, an important step to
ensure the self-consistency of the final result.
\item
An emission measure analysis was performed and element abundances (relative to iron) were determined,
using both line-by-line inverse emissivity curves
and the Markov-Chain Monte Carlo method of \citet{kd98}, 
as described in more detail in Sect.~\ref{sec:em}.
\item
Iron abundance was derived by comparison of the observed continuum with
predictions obtained by assuming $Fe/H$ ratios different from the solar ratio.
\item
Temperature and density diagnostics were inferred from He-like ion line ratios,
based both on the {\sc chianti} line emissivities and on theoretical 
calculations by \citet*{pmd01}
and taking into account the final EMD. Density
diagnostics using sensitive \ion{Fe}{21} line ratios were also
investigated.
\end{enumerate}

Each step of our analysis required considerable care
to ensure self-consistency and robustness of
the results and to understand the limits of our spectroscopic 
approach. In the following we describe
our most pertinent findings; further details are provided in the
Appendix.

\begin{figure*}[!ht]
   \centering
   \resizebox{0.8\hsize}{!}{
   \plotone{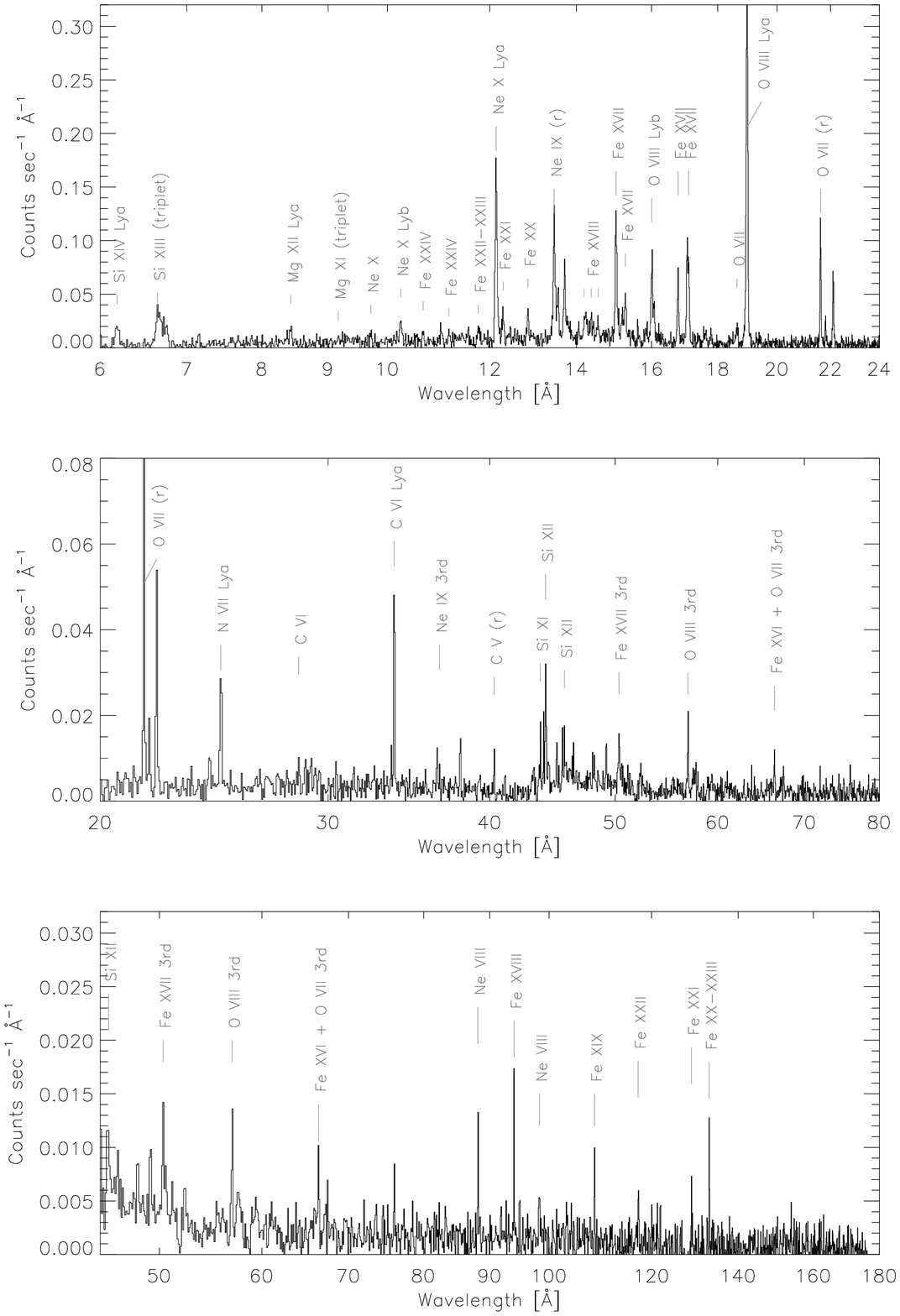}
   }
\caption{
AD Leo spectrum, shown in three overlapping wavelength ranges,
with line identifications (for the strongest lines) based on the {\sc chianti}
line list. Different Y-axis ranges have been used in the three panels.
The bin size is 0.0125\,\AA~in the first wavelength range, while
a rebinning by factors 4 and 8 has been performed in the second and
third range, respectively. Note that the X-axis scale is logarithmic.
\label{fig:sp}}
\end{figure*}

The observed spectrum, with identifications of the emitting ion for the most
prominent features, is shown in Fig.~\ref{fig:sp}. 
In total, 120 lines have been
identified, including some lines from the overlapping 3rd order spectrum,
as reported in Table~\ref{tab:lines}.
Iron lines from ionization stages XVI through XXIV have been clearly 
detected, as well as lines from C, N, O, the other
most abundant $\alpha$ elements (Ne, Mg, Si, S, Ni), and one Al line.
Overall, the range of temperatures probed by the observed lines
extends from $T \sim 10^{5.8}$\,K (with the \ion{Ne}{8} lines) to 
$T \sim 10^{7.5}$\,K (with the \ion{Si}{14}, \ion{Fe}{23}, and 
\ion{Fe}{24} lines). 

After a careful examination of the identified lines, we have selected 50
of them for the emission measure analysis, giving preference to lines
that have high S/N ratios, are not affected by blending and are not density dependent
(see Sect.~\ref{sec:em}).  Noteworthy among the discarded lines
is the \ion{Ne}{10} Ly$\alpha$ doublet ($\lambda
12.12,12.13$\,\AA), blended with an \ion{Fe}{17} line that provides
about 5\% of the measured total flux: according to the final EMD and
element abundances this spectral feature appears to be suppressed,
in the sense that the measured \ion{Ne}{10} Ly$\alpha$/Ly$\beta$ and
Ly$\alpha$/Ly$\gamma$ ratios are a factor $\ga 2$ lower than expected. 
This result cannot be attributed to an underestimated contribution from the

\clearpage

\clearpage

\begin{table*}[!h]
\centering
\caption{Line identifications and fluxes in the AD Leo spectrum.}
\label{tab:lines}
\small
\begin{tabular}[h]{@{\hspace{4mm}}l@{\hspace{4mm}}rrllcr@{\hspace{1mm}}c@{\hspace{1mm}}lc}
\hline
\hspace{-2mm}$Label^{a}$ & $\lambda_{\mathrm{obs}}^{b}$\hspace{-3mm} & $\lambda_{\mathrm{pred}}^{b}$ & \multicolumn{1}{c}{Ion} & \multicolumn{1}{l}{Transition $(upper \rightarrow lower)$} & $\log T_{\mathrm{max}}^{c}$ & $(F\,$ & $\pm$ & $\sigma)^{d}$ & $EM^{e}$ \\
\hline
 ~\,\,  1a &   6.18 &   6.18 & \ion{Si}{    14} &                                            $\;2p^{}\;^{2}P_{3/2}\; \rightarrow \;1s^{}\;^{2}S_{1/2}$ & 7.20 &       50 & $\pm$ &  11 &  Y \\
 ~\,\,  1b & $\cdots$ &   6.19 & \ion{Si}{    14} &                                            $\;2p^{}\;^{2}P_{1/2}\; \rightarrow \;1s^{}\;^{2}S_{1/2}$ & 7.20 & $\cdots$ &       &     &    \\
  ~\,\,  2 &   6.64 &   6.65 & \ion{Si}{    13} &                                           $\;1s\;2p^{}\;^{1}P_{1}\; \rightarrow \;1s^{2}\;^{1}S_{0}$ & 7.00 &       69 & $\pm$ &  14 &  Y \\
 ~\,\,  3a &   6.68 &   6.69 & \ion{Si}{    13} &                                           $\;1s\;2p^{}\;^{3}P_{2}\; \rightarrow \;1s^{2}\;^{1}S_{0}$ & 6.95 &       48 & $\pm$ &  13 &    \\
 ~\,\,  3b & $\cdots$ &   6.69 & \ion{Si}{    13} &                                           $\;1s\;2p^{}\;^{3}P_{1}\; \rightarrow \;1s^{2}\;^{1}S_{0}$ & 6.95 & $\cdots$ &       &     &    \\
 ~\,\,  4a &   6.72 &   6.72 & \ion{Si}{    12} &                   $\;1s\;2s\;(^{3}P)\;2p^{}\;^{2}P_{3/2}\; \rightarrow \;1s^{2}\;2s^{}\;^{2}S_{1/2}$ & 6.95 &       19 & $\pm$ &  11 &    \\
 ~\,\,  4b & $\cdots$ &   6.72 & \ion{Si}{    12} &                   $\;1s\;2s\;(^{3}P)\;2p^{}\;^{2}P_{1/2}\; \rightarrow \;1s^{2}\;2s^{}\;^{2}S_{1/2}$ & 6.95 & $\cdots$ &       &     &    \\
  ~\,\,  5 &   6.75 &   6.74 & \ion{Si}{    13} &                                           $\;1s\;2s^{}\;^{3}S_{1}\; \rightarrow \;1s^{2}\;^{1}S_{0}$ & 7.00 &       32 & $\pm$ &  11 &    \\
 ~\,\,  6a &   7.15 &   7.17 & \ion{Al}{    13} &                                            $\;2p^{}\;^{2}P_{3/2}\; \rightarrow \;1s^{}\;^{2}S_{1/2}$ & 7.10 &       14 & $\pm$ &   8 &    \\
% ~\,\,  6b & $\cdots$ &   7.17 & \ion{Fe}{    24} &                            $\;1s^{2}\;5p^{}\;^{2}P_{1/2}\; \rightarrow \;1s^{2}\;2s^{}\;^{2}S_{1/2}$ & 7.30 & $\cdots$ &       &     &    \\
 ~\,\,  6b & $\cdots$ &   7.18 & \ion{Al}{    13} &                                            $\;2p^{}\;^{2}P_{1/2}\; \rightarrow \;1s^{}\;^{2}S_{1/2}$ & 7.10 & $\cdots$ &       &     &    \\
 ~\,\,  7a &   8.42 &   8.42 & \ion{Mg}{    12} &                                            $\;2p^{}\;^{2}P_{3/2}\; \rightarrow \;1s^{}\;^{2}S_{1/2}$ & 7.00 &       37 & $\pm$ &  10 &  Y \\
 ~\,\,  7b & $\cdots$ &   8.42 & \ion{Mg}{    12} &                                            $\;2p^{}\;^{2}P_{1/2}\; \rightarrow \;1s^{}\;^{2}S_{1/2}$ & 7.00 & $\cdots$ &       &     &    \\
  ~\,\,  8 &   8.83 &   8.81 & \ion{Fe}{    23} &                                        $\;2s\;4d^{}\;^{1}D_{2}\; \rightarrow \;2s\;2p^{}\;^{1}P_{1}$ & 7.20 &       15 & $\pm$ &   8 &    \\
  ~\,\,  9 &   9.16 &   9.17 & \ion{Mg}{    11} &                                           $\;1s\;2p^{}\;^{1}P_{1}\; \rightarrow \;1s^{2}\;^{1}S_{0}$ & 6.80 &       22 & $\pm$ &   9 &  Y \\
    \, 10a &   9.23 &   9.23 & \ion{Mg}{    11} &                                           $\;1s\;2p^{}\;^{3}P_{2}\; \rightarrow \;1s^{2}\;^{1}S_{0}$ & 6.80 &       15 & $\pm$ &   8 &    \\
    \, 10b & $\cdots$ &   9.23 & \ion{Mg}{    11} &                                           $\;1s\;2p^{}\;^{3}P_{1}\; \rightarrow \;1s^{2}\;^{1}S_{0}$ & 6.80 & $\cdots$ &       &     &    \\
     \, 11 &   9.31 &   9.31 & \ion{Mg}{    11} &                                           $\;1s\;2s^{}\;^{3}S_{1}\; \rightarrow \;1s^{2}\;^{1}S_{0}$ & 6.80 &       24 & $\pm$ &   9 &    \\
    \, 12a &   9.71 &   9.71 & \ion{Ne}{    10} &                                            $\;4p^{}\;^{2}P_{3/2}\; \rightarrow \;1s^{}\;^{2}S_{1/2}$ & 6.80 &       25 & $\pm$ &   9 &  Y \\
    \, 12b & $\cdots$ &   9.71 & \ion{Ne}{    10} &                                            $\;4p^{}\;^{2}P_{1/2}\; \rightarrow \;1s^{}\;^{2}S_{1/2}$ & 6.80 & $\cdots$ &       &     &    \\
    \, 13a &  10.24 &  10.24 & \ion{Ne}{    10} &                                            $\;3p^{}\;^{2}P_{3/2}\; \rightarrow \;1s^{}\;^{2}S_{1/2}$ & 6.80 &       66 & $\pm$ &  12 &  Y \\
    \, 13b & $\cdots$ &  10.24 & \ion{Ne}{    10} &                                            $\;3p^{}\;^{2}P_{1/2}\; \rightarrow \;1s^{}\;^{2}S_{1/2}$ & 6.80 & $\cdots$ &       &     &    \\
    \, 14a &  10.66 &  10.62 & \ion{Fe}{    24} &                            $\;1s^{2}\;3p^{}\;^{2}P_{3/2}\; \rightarrow \;1s^{2}\;2s^{}\;^{2}S_{1/2}$ & 7.30 &       32 & $\pm$ &  10 &  Y \\
    \, 14b & $\cdots$ &  10.66 & \ion{Fe}{    24} &                            $\;1s^{2}\;3p^{}\;^{2}P_{1/2}\; \rightarrow \;1s^{2}\;2s^{}\;^{2}S_{1/2}$ & 7.30 & $\cdots$ &       &     &    \\
    \, 15a &  11.00 &  10.98 & \ion{Fe}{    23} &                                           $\;2s\;3p^{}\;^{1}P_{1}\; \rightarrow \;2s^{2}\;^{1}S_{0}$ & 7.20 &       33 & $\pm$ &  10 &    \\
    \, 15b & $\cdots$ &  11.00 & \ion{Ne}{     9} &                                           $\;1s\;4p^{}\;^{1}P_{1}\; \rightarrow \;1s^{2}\;^{1}S_{0}$ & 6.60 & $\cdots$ &       &     &    \\
    \, 16a &  11.03 &  11.02 & \ion{Fe}{    23} &                                           $\;2s\;3p^{}\;^{3}P_{1}\; \rightarrow \;2s^{2}\;^{1}S_{0}$ & 7.20 &       12 & $\pm$ &   9 &    \\
    \, 16b & $\cdots$ &  11.02 & \ion{Fe}{    17} &                                   $\;2s\;2p^{6}\;4p^{}\;^{1}P_{1}\; \rightarrow \;2p^{6}\;^{1}S_{0}$ & 6.80 & $\cdots$ &       &     &    \\
    \, 16c & $\cdots$ &  11.03 & \ion{Fe}{    24} &                            $\;1s^{2}\;3d^{}\;^{2}D_{3/2}\; \rightarrow \;1s^{2}\;2p^{}\;^{2}P_{1/2}$ & 7.30 & $\cdots$ &       &     &    \\
     \, 17 &  11.16 &  11.17 & \ion{Fe}{    24} &                            $\;1s^{2}\;3d^{}\;^{2}D_{5/2}\; \rightarrow \;1s^{2}\;2p^{}\;^{2}P_{3/2}$ & 7.30 &       33 & $\pm$ &  10 &  Y \\
     \, 18 &  11.24 &  11.26 & \ion{Fe}{    24} &                            $\;1s^{2}\;3s^{}\;^{2}S_{1/2}\; \rightarrow \;1s^{2}\;2p^{}\;^{2}P_{1/2}$ & 7.30 &       11 & $\pm$ &   8 &  Y \\
    \, 19a &  11.42 &  11.43 & \ion{Fe}{    24} &                            $\;1s^{2}\;3s^{}\;^{2}S_{1/2}\; \rightarrow \;1s^{2}\;2p^{}\;^{2}P_{3/2}$ & 7.30 &       36 & $\pm$ &  10 &    \\
    \, 19b & $\cdots$ &  11.44 & \ion{Fe}{    23} &                                        $\;2s\;3d^{}\;^{3}D_{3}\; \rightarrow \;2s\;2p^{}\;^{3}P_{2}$ & 7.15 & $\cdots$ &       &     &    \\
    \, 19c & $\cdots$ &  11.44 & \ion{Fe}{    22} &                   $\;2s\;2p\;(^{3}P)\;3p^{}\;^{4}S_{3/2}\; \rightarrow \;2s^{2}\;2p^{}\;^{2}P_{1/2}$ & 7.10 & $\cdots$ &       &     &    \\
     \, 20 &  11.54 &  11.55 & \ion{Ne}{     9} &                                           $\;1s\;3p^{}\;^{1}P_{1}\; \rightarrow \;1s^{2}\;^{1}S_{0}$ & 6.60 &       38 & $\pm$ &  10 &  Y \\
     \, 21 &  11.75 &  11.74 & \ion{Fe}{    23} &                                        $\;2s\;3d^{}\;^{1}D_{2}\; \rightarrow \;2s\;2p^{}\;^{1}P_{1}$ & 7.20 &       38 & $\pm$ &  11 &  Y \\
     \, 22 &  11.79 &  11.77 & \ion{Fe}{    22} &                   $\;2s^{2}\;(^{1}S)\;3d^{}\;^{2}D_{3/2}\; \rightarrow \;2s^{2}\;2p^{}\;^{2}P_{1/2}$ & 7.10 &       18 & $\pm$ &  10 &  Y \\
    \, 23a &  12.14 &  12.12 & \ion{Fe}{    17} &                                       $\;2p^{5}\;4d^{}\;^{1}P_{1}\; \rightarrow \;2p^{6}\;^{1}S_{0}$ & 6.80 &      436 & $\pm$ &  24 &  Y \\
    \, 23b & $\cdots$ &  12.13 & \ion{Ne}{    10} &                                            $\;2p^{}\;^{2}P_{3/2}\; \rightarrow \;1s^{}\;^{2}S_{1/2}$ & 6.80 & $\cdots$ &       &     &    \\
    \, 23c & $\cdots$ &  12.14 & \ion{Ne}{    10} &                                            $\;2p^{}\;^{2}P_{1/2}\; \rightarrow \;1s^{}\;^{2}S_{1/2}$ & 6.80 & $\cdots$ &       &     &    \\
     \, 24 &  12.26 &  12.26 & \ion{Fe}{    17} &                                       $\;2p^{5}\;4d^{}\;^{3}D_{1}\; \rightarrow \;2p^{6}\;^{1}S_{0}$ & 6.80 &       16 & $\pm$ &  11 &    \\
\hline
\end{tabular}
\normalsize
\end{table*}
\begin{table*}[!h]
\centering
\parbox{14cm}{
\bf{Table~\ref{tab:lines}.}\small\rm~Continued
\vspace{5mm}
}
\small
\begin{tabular}[h]{@{\hspace{4mm}}l@{\hspace{4mm}}rrllcr@{\hspace{1mm}}c@{\hspace{1mm}}lc}
\hline
\hspace{-2mm}$Label^{a}$ & $\lambda_{\mathrm{obs}}^{b}$\hspace{-3mm} & $\lambda_{\mathrm{pred}}^{b}$ & \multicolumn{1}{c}{Ion} & \multicolumn{1}{l}{Transition $(upper \rightarrow lower)$} & $\log T_{\mathrm{max}}^{c}$ & $(F\,$ & $\pm$ & $\sigma)^{d}$ & $EM^{e}$ \\
\hline
     \, 25 &  12.29 &  12.28 & \ion{Fe}{    21} &                           $\;2s^{2}\;2p\;3d^{}\;^{3}D_{1}\; \rightarrow \;2s^{2}\;2p^{2}\;^{3}P_{0}$ & 7.00 &       67 & $\pm$ &  13 &  Y \\
     \, 26 &  12.38 &  12.38 & \ion{Fe}{    21} &                           $\;2s^{2}\;2p\;3d^{}\;^{3}D_{3}\; \rightarrow \;2s^{2}\;2p^{2}\;^{3}P_{2}$ & 7.00 &       14 & $\pm$ &   9 &    \\
    \, 27a &  12.41 &  12.40 & \ion{Fe}{    21} &                           $\;2s^{2}\;2p\;3d^{}\;^{3}D_{1}\; \rightarrow \;2s^{2}\;2p^{2}\;^{3}P_{1}$ & 7.00 &       14 & $\pm$ &  10 &  Y \\
    \, 27b & $\cdots$ &  12.42 & \ion{Fe}{    21} &                           $\;2s^{2}\;2p\;3d^{}\;^{3}D_{2}\; \rightarrow \;2s^{2}\;2p^{2}\;^{3}P_{1}$ & 7.00 & $\cdots$ &       &     &    \\
     \, 28 &  12.45 &  12.44 & \ion{Ni}{    19} &                                       $\;2p^{5}\;3d^{}\;^{1}P_{1}\; \rightarrow \;2p^{6}\;^{1}S_{0}$ & 6.90 &       30 & $\pm$ &  10 &  Y \\
     \, 29 &  12.54 &  12.52 & \ion{Fe}{    21} &                           $\;2s^{2}\;2p\;3d^{}\;^{3}F_{3}\; \rightarrow \;2s^{2}\;2p^{2}\;^{3}P_{2}$ & 7.00 &       27 & $\pm$ &  10 &    \\
     \, 30 &  12.60 &  12.59 & \ion{Fe}{    21} &                           $\;2s^{2}\;2p\;3d^{}\;^{3}F_{2}\; \rightarrow \;2s^{2}\;2p^{2}\;^{3}P_{2}$ & 7.00 &       27 & $\pm$ &  10 &    \\
    \, 31a &  12.84 &  12.82 & \ion{Fe}{    20} &          $\;2s^{2}\;2p^{2}\;(^{3}P)\;3d^{}\;^{4}P_{5/2}\; \rightarrow \;2s^{2}\;2p^{3}\;^{4}S_{3/2}$ & 7.00 &       83 & $\pm$ &  13 &  Y \\
    \, 31b & $\cdots$ &  12.83 & \ion{Fe}{    20} &          $\;2s^{2}\;2p^{2}\;(^{3}P)\;3d^{}\;^{4}P_{1/2}\; \rightarrow \;2s^{2}\;2p^{3}\;^{4}S_{3/2}$ & 7.00 & $\cdots$ &       &     &    \\
    \, 31c & $\cdots$ &  12.83 & \ion{Fe}{    20} &          $\;2s^{2}\;2p^{2}\;(^{3}P)\;3d^{}\;^{4}P_{3/2}\; \rightarrow \;2s^{2}\;2p^{3}\;^{4}S_{3/2}$ & 7.00 & $\cdots$ &       &     &    \\
     \, 32 &  13.41 &  13.42 & \ion{Fe}{    19} &                      $\;2p^{3}\;(^{2}D)\;3d^{}\;^{1}F_{3}\; \rightarrow \;2s^{2}\;2p^{4}\;^{3}P_{2}$ & 6.90 &       26 & $\pm$ &  12 &    \\
    \, 33a &  13.45 &  13.45 & \ion{Ne}{     9} &                                           $\;1s\;2p^{}\;^{1}P_{1}\; \rightarrow \;1s^{2}\;^{1}S_{0}$ & 6.60 &      281 & $\pm$ &  22 &  Y \\
    \, 33b & $\cdots$ &  13.46 & \ion{Fe}{    19} &                      $\;2p^{3}\;(^{2}D)\;3d^{}\;^{3}D_{1}\; \rightarrow \;2s^{2}\;2p^{4}\;^{3}P_{0}$ & 6.90 & $\cdots$ &       &     &    \\
    \, 33c & $\cdots$ &  13.46 & \ion{Fe}{    19} &                      $\;2p^{3}\;(^{2}D)\;3d^{}\;^{3}S_{1}\; \rightarrow \;2s^{2}\;2p^{4}\;^{3}P_{2}$ & 6.90 & $\cdots$ &       &     &    \\
     \, 34 &  13.49 &  13.50 & \ion{Fe}{    19} &                      $\;2p^{3}\;(^{2}P)\;3d^{}\;^{1}D_{2}\; \rightarrow \;2s^{2}\;2p^{4}\;^{3}P_{2}$ & 6.90 &       48 & $\pm$ &  17 &    \\
    \, 35a &  13.52 &  13.51 & \ion{Fe}{    21} &                      $\;2s\;2p^{2}\;(^{4}P)\;3s^{}\;^{3}P_{0}\; \rightarrow \;2s\;2p^{3}\;^{3}D_{1}$ & 7.00 &       15 & $\pm$ &  15 &    \\
    \, 35b & $\cdots$ &  13.52 & \ion{Fe}{    19} &                      $\;2p^{3}\;(^{2}D)\;3d^{}\;^{3}D_{3}\; \rightarrow \;2s^{2}\;2p^{4}\;^{3}P_{2}$ & 6.90 & $\cdots$ &       &     &    \\
    \, 36a &  13.55 &  13.53 & \ion{Fe}{    19} &                      $\;2p^{3}\;(^{2}D)\;3d^{}\;^{3}D_{2}\; \rightarrow \;2s^{2}\;2p^{4}\;^{3}P_{2}$ & 6.90 &      120 & $\pm$ &  16 &    \\
    \, 36b & $\cdots$ &  13.55 & \ion{Ne}{     9} &                                           $\;1s\;2p^{}\;^{3}P_{2}\; \rightarrow \;1s^{2}\;^{1}S_{0}$ & 6.55 & $\cdots$ &       &     &    \\
    \, 36c & $\cdots$ &  13.55 & \ion{Fe}{    19} &                      $\;2p^{3}\;(^{2}P)\;3d^{}\;^{3}D_{2}\; \rightarrow \;2s^{2}\;2p^{4}\;^{3}P_{1}$ & 6.90 & $\cdots$ &       &     &    \\
    \, 36d & $\cdots$ &  13.55 & \ion{Ne}{     9} &                                           $\;1s\;2p^{}\;^{3}P_{1}\; \rightarrow \;1s^{2}\;^{1}S_{0}$ & 6.55 & $\cdots$ &       &     &    \\
    \, 37a &  13.67 &  13.65 & \ion{Fe}{    19} &                      $\;2p^{3}\;(^{2}D)\;3d^{}\;^{3}F_{3}\; \rightarrow \;2s^{2}\;2p^{4}\;^{3}P_{2}$ & 6.90 &       30 & $\pm$ &  13 &    \\
    \, 37b & $\cdots$ &  13.66 & \ion{Fe}{    19} &                      $\;2p^{3}\;(^{2}D)\;3d^{}\;^{1}P_{1}\; \rightarrow \;2s^{2}\;2p^{4}\;^{3}P_{2}$ & 6.90 & $\cdots$ &       &     &    \\
     \, 38 &  13.71 &  13.70 & \ion{Ne}{     9} &                                           $\;1s\;2s^{}\;^{3}S_{1}\; \rightarrow \;1s^{2}\;^{1}S_{0}$ & 6.60 &      182 & $\pm$ &  18 &    \\
    \, 39a &  13.74 &  13.74 & \ion{Fe}{    19} &                      $\;2p^{3}\;(^{2}D)\;3d^{}\;^{3}P_{2}\; \rightarrow \;2s^{2}\;2p^{4}\;^{3}P_{1}$ & 6.90 &  $\sim 5$&  &  &    \\
    \, 39b & $\cdots$ &  13.74 & \ion{Fe}{    19} &                      $\;2p^{3}\;(^{2}D)\;3d^{}\;^{1}F_{3}\; \rightarrow \;2s^{2}\;2p^{4}\;^{1}D_{2}$ & 6.90 & $\cdots$ &       &     &    \\
    \, 39c & $\cdots$ &  13.74 & \ion{Fe}{    20} &          $\;2s^{2}\;2p^{2}\;(^{3}P)\;3s^{}\;^{4}P_{5/2}\; \rightarrow \;2s^{2}\;2p^{3}\;^{4}S_{3/2}$ & 7.00 & $\cdots$ &       &     &    \\
    \, 40a &  13.78 &  13.78 & \ion{Fe}{    19} &                      $\;2p^{3}\;(^{2}D)\;3d^{}\;^{3}S_{1}\; \rightarrow \;2s^{2}\;2p^{4}\;^{1}D_{2}$ & 6.90 &       54 & $\pm$ &  12 &    \\
    \, 40b & $\cdots$ &  13.78 & \ion{Ni}{    19} &                                       $\;2p^{5}\;3s^{}\;^{1}P_{1}\; \rightarrow \;2p^{6}\;^{1}S_{0}$ & 6.80 & $\cdots$ &       &     &    \\
    \, 40c & $\cdots$ &  13.80 & \ion{Fe}{    19} &                      $\;2p^{3}\;(^{4}S)\;3d^{}\;^{3}D_{3}\; \rightarrow \;2s^{2}\;2p^{4}\;^{3}P_{2}$ & 6.90 & $\cdots$ &       &     &    \\
     \, 41 &  13.84 &  13.82 & \ion{Fe}{    17} &                                   $\;2s\;2p^{6}\;3p^{}\;^{1}P_{1}\; \rightarrow \;2p^{6}\;^{1}S_{0}$ & 6.80 &       41 & $\pm$ &  11 &    \\
    \, 42a &  14.06 &  14.04 & \ion{Ni}{    19} &                                       $\;2p^{5}\;3s^{}\;^{3}P_{1}\; \rightarrow \;2p^{6}\;^{1}S_{0}$ & 6.80 &       50 & $\pm$ &  11 &  Y \\
    \, 42b & $\cdots$ &  14.08 & \ion{Ni}{    19} &                                       $\;2p^{5}\;3s^{}\;^{3}P_{2}\; \rightarrow \;2p^{6}\;^{1}S_{0}$ & 6.80 & $\cdots$ &       &     &    \\
    \, 43a &  14.19 &  14.20 & \ion{Fe}{    18} &                  $\;2p^{4}\;(^{1}D)\;3d^{}\;^{2}D_{5/2}\; \rightarrow \;2s^{2}\;2p^{5}\;^{2}P_{3/2}$ & 6.90 &       53 & $\pm$ &  13 &   \\
    \, 43b & $\cdots$ &  14.21 & \ion{Fe}{    18} &                  $\;2p^{4}\;(^{1}D)\;3d^{}\;^{2}P_{3/2}\; \rightarrow \;2s^{2}\;2p^{5}\;^{2}P_{3/2}$ & 6.90 & $\cdots$ &       &     &    \\
     \, 44 &  14.23 &  14.26 & \ion{Fe}{    18} &                  $\;2p^{4}\;(^{1}D)\;3d^{}\;^{2}S_{1/2}\; \rightarrow \;2s^{2}\;2p^{5}\;^{2}P_{3/2}$ & 6.90 &       64 & $\pm$ &  14 &    \\
    \, 45a &  14.27 &  14.27 & \ion{Fe}{    18} &                  $\;2p^{4}\;(^{1}D)\;3d^{}\;^{2}F_{5/2}\; \rightarrow \;2s^{2}\;2p^{5}\;^{2}P_{3/2}$ & 6.90 &       29 & $\pm$ &  12 &    \\
    \, 45b & $\cdots$ &  14.27 & \ion{Fe}{    20} &                  $\;2s\;2p^{3}\;(^{5}S)\;3s^{}\;^{4}S_{3/2}\; \rightarrow \;2s\;2p^{4}\;^{4}P_{5/2}$ & 7.00 & $\cdots$ &       &     &    \\
     \, 46 &  14.32 &  14.34 & \ion{Fe}{    18} &                  $\;2p^{4}\;(^{1}D)\;3d^{}\;^{2}P_{1/2}\; \rightarrow \;2s^{2}\;2p^{5}\;^{2}P_{1/2}$ & 6.90 &       27 & $\pm$ &  11 &    \\
\hline
\end{tabular}
\normalsize
\end{table*}
\begin{table*}[!h]
\centering
\parbox{14cm}{
\bf{Table~\ref{tab:lines}.}\small\rm~Continued
\vspace{5mm}
}
\small
\begin{tabular}[h]{@{\hspace{4mm}}l@{\hspace{4mm}}rrllcr@{\hspace{1mm}}c@{\hspace{1mm}}lc}
\hline
\hspace{-2mm}$Label^{a}$ & $\lambda_{\mathrm{obs}}^{b}$\hspace{-3mm} & $\lambda_{\mathrm{pred}}^{b}$ & \multicolumn{1}{c}{Ion} & \multicolumn{1}{l}{Transition $(upper \rightarrow lower)$} & $\log T_{\mathrm{max}}^{c}$ & $(F\,$ & $\pm$ & $\sigma)^{d}$ & $EM^{e}$ \\
\hline
    \, 47a &  14.37 &  14.36 & \ion{Fe}{    18} &                  $\;2p^{4}\;(^{1}D)\;3d^{}\;^{2}D_{3/2}\; \rightarrow \;2s^{2}\;2p^{5}\;^{2}P_{1/2}$ & 6.90 &       64 & $\pm$ &  13 &  Y \\
    \, 47b & $\cdots$ &  14.37 & \ion{Fe}{    18} &                  $\;2p^{4}\;(^{3}P)\;3d^{}\;^{2}D_{5/2}\; \rightarrow \;2s^{2}\;2p^{5}\;^{2}P_{3/2}$ & 6.90 & $\cdots$ &       &     &    \\
    \, 48a &  14.43 &  14.41 & \ion{Fe}{    20} &                  $\;2s\;2p^{3}\;(^{5}S)\;3s^{}\;^{4}S_{3/2}\; \rightarrow \;2s\;2p^{4}\;^{4}P_{3/2}$ & 7.00 &       46 & $\pm$ &  11 &    \\
    \, 48b & $\cdots$ &  14.42 & \ion{Fe}{    18} &                  $\;2p^{4}\;(^{1}D)\;3d^{}\;^{2}P_{3/2}\; \rightarrow \;2s^{2}\;2p^{5}\;^{2}P_{1/2}$ & 6.90 & $\cdots$ &       &     &    \\
    \, 49a &  14.51 &  14.48 & \ion{Fe}{    18} &                  $\;2p^{4}\;(^{3}P)\;3d^{}\;^{4}F_{5/2}\; \rightarrow \;2s^{2}\;2p^{5}\;^{2}P_{3/2}$ & 6.90 &       30 & $\pm$ &  11 &    \\
    \, 49b & $\cdots$ &  14.51 & \ion{Fe}{    20} &              $\;2s^{2}\;2p^{2}\;(^{1}D)\;3p^{}\;^{2}P_{3/2}\; \rightarrow \;2s\;2p^{4}\;^{4}P_{1/2}$ & 7.00 & $\cdots$ &       &     &    \\
    \, 50a &  14.55 &  14.53 & \ion{Fe}{    18} &                  $\;2p^{4}\;(^{3}P)\;3d^{}\;^{2}F_{5/2}\; \rightarrow \;2s^{2}\;2p^{5}\;^{2}P_{3/2}$ & 6.90 &       57 & $\pm$ &  12 &  Y \\
    \, 50b & $\cdots$ &  14.55 & \ion{Fe}{    18} &                  $\;2p^{4}\;(^{3}P)\;3d^{}\;^{4}P_{3/2}\; \rightarrow \;2s^{2}\;2p^{5}\;^{2}P_{3/2}$ & 6.90 & $\cdots$ &       &     &    \\
    \, 50c & $\cdots$ &  14.58 & \ion{Fe}{    18} &                  $\;2p^{4}\;(^{3}P)\;3d^{}\;^{4}P_{1/2}\; \rightarrow \;2s^{2}\;2p^{5}\;^{2}P_{3/2}$ & 6.90 & $\cdots$ &       &     &    \\
     \, 51 &  15.02 &  15.02 & \ion{Fe}{    17} &                                       $\;2p^{5}\;3d^{}\;^{1}P_{1}\; \rightarrow \;2p^{6}\;^{1}S_{0}$ & 6.75 &      330 & $\pm$ &  22 &  Y \\
     \, 52 &  15.09 &  15.08 & \ion{Fe}{    19} &                      $\;2s\;2p^{4}\;(^{4}P)\;3s^{}\;^{3}P_{2}\; \rightarrow \;2s\;2p^{5}\;^{3}P_{2}$ & 6.90 &       47 & $\pm$ &  13 &    \\
    \, 53a &  15.16 &  15.18 & \ion{O }{     8} &                                            $\;4p^{}\;^{2}P_{3/2}\; \rightarrow \;1s^{}\;^{2}S_{1/2}$ & 6.50 &       29 & $\pm$ &  12 &    \\
    \, 53b & $\cdots$ &  15.18 & \ion{O }{     8} &                                            $\;4p^{}\;^{2}P_{1/2}\; \rightarrow \;1s^{}\;^{2}S_{1/2}$ & 6.50 & $\cdots$ &       &     &    \\
     \, 54 &  15.19 &  15.20 & \ion{Fe}{    19} &                      $\;2p^{3}\;(^{4}S)\;3s^{}\;^{5}S_{2}\; \rightarrow \;2s^{2}\;2p^{4}\;^{3}P_{2}$ & 6.90 &       72 & $\pm$ &  14 &    \\
     \, 55 &  15.27 &  15.26 & \ion{Fe}{    17} &                                       $\;2p^{5}\;3d^{}\;^{3}D_{1}\; \rightarrow \;2p^{6}\;^{1}S_{0}$ & 6.75 &      133 & $\pm$ &  15 &  Y \\
     \, 56 &  15.44 &  15.45 & \ion{Fe}{    17} &                                       $\;2p^{5}\;3d^{}\;^{3}P_{1}\; \rightarrow \;2p^{6}\;^{1}S_{0}$ & 6.70 &       29 & $\pm$ &  10 &    \\
     \, 57 &  15.49 &  15.50 & \ion{Fe}{    18} &                  $\;2s\;2p^{5}\;(^{1}P)\;3s^{}\;^{2}P_{3/2}\; \rightarrow \;2s\;2p^{6}\;^{2}S_{1/2}$ & 6.90 &       19 & $\pm$ &   9 &    \\
     \, 58 &  15.62 &  15.63 & \ion{Fe}{    18} &                  $\;2p^{4}\;(^{1}D)\;3s^{}\;^{2}D_{5/2}\; \rightarrow \;2s^{2}\;2p^{5}\;^{2}P_{3/2}$ & 6.90 &       31 & $\pm$ &  10 &   \\
     \, 59 &  15.66 &  15.63 & \ion{Fe}{    20} &              $\;2s^{2}\;2p^{2}\;(^{1}D)\;3p^{}\;^{2}P_{3/2}\; \rightarrow \;2s\;2p^{4}\;^{2}P_{1/2}$ & 7.00 &       16 & $\pm$ &   9 &    \\
     \, 60 &  15.78 &  15.77 & \ion{Fe}{    18} &                  $\;2p^{4}\;(^{3}P)\;3s^{}\;^{2}P_{1/2}\; \rightarrow \;2s^{2}\;2p^{5}\;^{2}P_{3/2}$ & 6.90 &       19 & $\pm$ &   9 &    \\
     \, 61 &  15.83 &  15.83 & \ion{Fe}{    18} &                  $\;2p^{4}\;(^{3}P)\;3s^{}\;^{4}P_{3/2}\; \rightarrow \;2s^{2}\;2p^{5}\;^{2}P_{3/2}$ & 6.90 &       23 & $\pm$ &  10 &   \\
    \, 62a &  15.88 &  15.84 & \ion{Fe}{    20} &              $\;2s^{2}\;2p^{2}\;(^{3}P)\;3p^{}\;^{2}P_{3/2}\; \rightarrow \;2s\;2p^{4}\;^{2}P_{3/2}$ & 7.00 &       42 & $\pm$ &  11 &    \\
    \, 62b & $\cdots$ &  15.85 & \ion{Fe}{    18} &                  $\;2p^{4}\;(^{3}P)\;3s^{}\;^{4}P_{1/2}\; \rightarrow \;2s^{2}\;2p^{5}\;^{2}P_{3/2}$ & 6.90 & $\cdots$ &       &     &    \\
    \, 62c & $\cdots$ &  15.87 & \ion{Fe}{    18} &                  $\;2p^{4}\;(^{1}D)\;3s^{}\;^{2}D_{3/2}\; \rightarrow \;2s^{2}\;2p^{5}\;^{2}P_{1/2}$ & 6.80 & $\cdots$ &       &     &    \\
    \, 63a &  16.01 &  16.00 & \ion{Fe}{    18} &                  $\;2p^{4}\;(^{3}P)\;3s^{}\;^{2}P_{3/2}\; \rightarrow \;2s^{2}\;2p^{5}\;^{2}P_{3/2}$ & 6.90 &      219 & $\pm$ &  19 &  Y \\
    \, 63b & $\cdots$ &  16.01 & \ion{O }{     8} &                                            $\;3p^{}\;^{2}P_{3/2}\; \rightarrow \;1s^{}\;^{2}S_{1/2}$ & 6.50 & $\cdots$ &       &     &    \\
    \, 63c & $\cdots$ &  16.01 & \ion{O }{     8} &                                            $\;3p^{}\;^{2}P_{1/2}\; \rightarrow \;1s^{}\;^{2}S_{1/2}$ & 6.50 & $\cdots$ &       &     &    \\
     \, 64 &  16.07 &  16.07 & \ion{Fe}{    18} &                  $\;2p^{4}\;(^{3}P)\;3s^{}\;^{4}P_{5/2}\; \rightarrow \;2s^{2}\;2p^{5}\;^{2}P_{3/2}$ & 6.90 &       94 & $\pm$ &  14 &   \\
    \, 65a &  16.20 &  16.17 & \ion{Fe}{    18} &                  $\;2s\;2p^{5}\;(^{3}P)\;3s^{}\;^{2}P_{3/2}\; \rightarrow \;2s\;2p^{6}\;^{2}S_{1/2}$ & 6.90 &       29 & $\pm$ &   9 &    \\
    \, 65b & $\cdots$ &  16.24 & \ion{Fe}{    17} &                                       $\;2p^{5}\;3p^{}\;^{3}P_{2}\; \rightarrow \;2p^{6}\;^{1}S_{0}$ & 6.70 & $\cdots$ &       &     &    \\
    \, 66a &  16.32 &  16.31 & \ion{Fe}{    18} &                  $\;2s\;2p^{5}\;(^{3}P)\;3s^{}\;^{4}P_{3/2}\; \rightarrow \;2s\;2p^{6}\;^{2}S_{1/2}$ & 6.90 &       22 & $\pm$ &   9 &    \\
    \, 66b & $\cdots$ &  16.34 & \ion{Fe}{    17} &                                       $\;2p^{5}\;3p^{}\;^{3}D_{2}\; \rightarrow \;2p^{6}\;^{1}S_{0}$ & 6.70 & $\cdots$ &       &     &    \\
     \, 67 &  16.77 &  16.78 & \ion{Fe}{    17} &                                       $\;2p^{5}\;3s^{}\;^{3}P_{1}\; \rightarrow \;2p^{6}\;^{1}S_{0}$ & 6.70 &      188 & $\pm$ &  17 &  Y \\
     \, 68 &  17.05 &  17.05 & \ion{Fe}{    17} &                                       $\;2p^{5}\;3s^{}\;^{1}P_{1}\; \rightarrow \;2p^{6}\;^{1}S_{0}$ & 6.70 &      225 & $\pm$ &  20 &  Y \\
     \, 69 &  17.09 &  17.10 & \ion{Fe}{    17} &                                       $\;2p^{5}\;3s^{}\;^{3}P_{2}\; \rightarrow \;2p^{6}\;^{1}S_{0}$ & 6.70 &      181 & $\pm$ &  19 &  Y \\
     \, 70 &  17.62 &  17.66 & \ion{Fe}{    18} &                      $\;2p^{4}\;(^{1}D)\;3p^{}\;^{2}P_{3/2}\; \rightarrow \;2s\;2p^{6}\;^{2}S_{1/2}$ & 6.90 &       38 & $\pm$ &  10 &    \\
     \, 71 &  17.79 &  17.77 & \ion{O }{     7} &                                           $\;1s\;4p^{}\;^{1}P_{1}\; \rightarrow \;1s^{2}\;^{1}S_{0}$ & 6.35 &       36 & $\pm$ &  10 &    \\
     \, 72 &  18.63 &  18.63 & \ion{O }{     7} &                                           $\;1s\;3p^{}\;^{1}P_{1}\; \rightarrow \;1s^{2}\;^{1}S_{0}$ & 6.35 &       51 & $\pm$ &  11 &  Y \\
\hline
\end{tabular}
\normalsize
\end{table*}
\begin{table*}[!h]
\centering
\parbox{14cm}{
\bf{Table~\ref{tab:lines}.}\small\rm~Continued
\vspace{5mm}
}
\small
\begin{tabular}[h]{@{\hspace{4mm}}l@{\hspace{4mm}}rrllcr@{\hspace{1mm}}c@{\hspace{1mm}}lc}
\hline
\hspace{-2mm}$Label^{a}$ & $\lambda_{\mathrm{obs}}^{b}$\hspace{-3mm} & $\lambda_{\mathrm{pred}}^{b}$ & \multicolumn{1}{c}{Ion} & \multicolumn{1}{l}{Transition $(upper \rightarrow lower)$} & $\log T_{\mathrm{max}}^{c}$ & $(F\,$ & $\pm$ & $\sigma)^{d}$ & $EM^{e}$ \\
\hline
    \, 73a &  18.96 &  18.97 & \ion{O }{     8} &                                            $\;2p^{}\;^{2}P_{3/2}\; \rightarrow \;1s^{}\;^{2}S_{1/2}$ & 6.50 &     1214 & $\pm$ &  38 &  Y \\
    \, 73b & $\cdots$ &  18.97 & \ion{O }{     8} &                                            $\;2p^{}\;^{2}P_{1/2}\; \rightarrow \;1s^{}\;^{2}S_{1/2}$ & 6.50 & $\cdots$ &       &     &    \\
    \, 74a &  20.90 &  20.91 & \ion{N }{     7} &                                            $\;3p^{}\;^{2}P_{3/2}\; \rightarrow \;1s^{}\;^{2}S_{1/2}$ & 6.35 &       21 & $\pm$ &   8 &  Y \\
    \, 74b & $\cdots$ &  20.91 & \ion{N }{     7} &                                            $\;3p^{}\;^{2}P_{1/2}\; \rightarrow \;1s^{}\;^{2}S_{1/2}$ & 6.35 & $\cdots$ &       &     &    \\
     \, 75 &  21.61 &  21.60 & \ion{O }{     7} &                                           $\;1s\;2p^{}\;^{1}P_{1}\; \rightarrow \;1s^{2}\;^{1}S_{0}$ & 6.30 &      283 & $\pm$ &  20 &  Y \\
    \, 76a &  21.81 &  21.80 & \ion{O }{     7} &                                           $\;1s\;2p^{}\;^{3}P_{2}\; \rightarrow \;1s^{2}\;^{1}S_{0}$ & 6.30 &       59 & $\pm$ &  11 &    \\
    \, 76b & $\cdots$ &  21.81 & \ion{O }{     7} &                                           $\;1s\;2p^{}\;^{3}P_{1}\; \rightarrow \;1s^{2}\;^{1}S_{0}$ & 6.30 & $\cdots$ &       &     &    \\
     \, 77 &  22.10 &  22.10 & \ion{O }{     7} &                                           $\;1s\;2s^{}\;^{3}S_{1}\; \rightarrow \;1s^{2}\;^{1}S_{0}$ & 6.30 &      185 & $\pm$ &  17 &    \\
    \, 78a &  24.26 &  24.20 & \ion{S }{    14} &                            $\;1s^{2}\;4d^{}\;^{2}D_{3/2}\; \rightarrow \;1s^{2}\;2p^{}\;^{2}P_{1/2}$ & 6.50 &       30 & $\pm$ &   9 &    \\
    \, 78b & $\cdots$ &  24.28 & \ion{S }{    14} &                            $\;1s^{2}\;4d^{}\;^{2}D_{5/2}\; \rightarrow \;1s^{2}\;2p^{}\;^{2}P_{3/2}$ & 6.50 & $\cdots$ &       &     &    \\
    \, 79a &  24.78 &  24.78 & \ion{N }{     7} &                                            $\;2p^{}\;^{2}P_{3/2}\; \rightarrow \;1s^{}\;^{2}S_{1/2}$ & 6.30 &      126 & $\pm$ &  15 &  Y \\
    \, 79b & $\cdots$ &  24.78 & \ion{N }{     7} &                                            $\;2p^{}\;^{2}P_{1/2}\; \rightarrow \;1s^{}\;^{2}S_{1/2}$ & 6.30 & $\cdots$ &       &     &    \\
    \, 80a &  28.46 &  28.47 & \ion{C }{     6} &                                            $\;3p^{}\;^{2}P_{3/2}\; \rightarrow \;1s^{}\;^{2}S_{1/2}$ & 6.20 &       31 & $\pm$ &   9 &  Y \\
    \, 80b & $\cdots$ &  28.47 & \ion{C }{     6} &                                            $\;3p^{}\;^{2}P_{1/2}\; \rightarrow \;1s^{}\;^{2}S_{1/2}$ & 6.20 & $\cdots$ &       &     &    \\
     \, 81 &  28.79 &  28.79 & \ion{N }{     6} &                                           $\;1s\;2p^{}\;^{1}P_{1}\; \rightarrow \;1s^{2}\;^{1}S_{0}$ & 6.20 &       38 & $\pm$ &  10 &  Y \\
     \, 82 &  28.92 &  &          No Id &                                                                                                  ... & &       17 & $\pm$ &   8 &    \\
     \, 83 &  29.10 &  29.08 & \ion{N }{     6} &                                           $\;1s\;2p^{}\;^{3}P_{1}\; \rightarrow \;1s^{2}\;^{1}S_{0}$ & 6.15 &       34 & $\pm$ &   9 &    \\
     \, 84 &  29.30 &  29.32 & \ion{S }{    15} &                                        $\;1s\;3d^{}\;^{1}D_{2}\; \rightarrow \;1s\;2p^{}\;^{1}P_{1}$ & 7.20 &       26 & $\pm$ &   9 &    \\
     \, 85 &  29.53 &  29.53 & \ion{N }{     6} &                                           $\;1s\;2s^{}\;^{3}S_{1}\; \rightarrow \;1s^{2}\;^{1}S_{0}$ & 6.15 &       25 & $\pm$ &   9 &  Y \\
     \, 86 &  33.56 &  33.55 & \ion{S }{    14} &                            $\;1s^{2}\;3s^{}\;^{2}S_{1/2}\; \rightarrow \;1s^{2}\;2p^{}\;^{2}P_{3/2}$ & 6.50 &       39 & $\pm$ &  10 &  Y \\
    \, 87a &  33.74 &  33.73 & \ion{C }{     6} &                                            $\;2p^{}\;^{2}P_{3/2}\; \rightarrow \;1s^{}\;^{2}S_{1/2}$ & 6.20 &      226 & $\pm$ &  18 &  Y \\
    \, 87b & $\cdots$ &  33.74 & \ion{C }{     6} &                                            $\;2p^{}\;^{2}P_{1/2}\; \rightarrow \;1s^{}\;^{2}S_{1/2}$ & 6.20 & $\cdots$ &       &     &    \\
     \, 88 &  36.40 &  & \multicolumn{2}{l}{\ion{Fe}{17} + \ion{Ne}{10} Ly$\alpha$ 3rd order} & &       44 & $\pm$ &  10 &    \\
    \, 89a &  36.57 &  36.56 & \ion{S }{    12} &                            $\;2s^{2}\;3d^{}\;^{2}D_{5/2}\; \rightarrow \;2s^{2}\;2p^{}\;^{2}P_{3/2}$ & 6.40 &       24 & $\pm$ &   9 &  Y \\
    \, 89b & $\cdots$ &  36.57 & \ion{S }{    12} &                            $\;2s^{2}\;3d^{}\;^{2}D_{3/2}\; \rightarrow \;2s^{2}\;2p^{}\;^{2}P_{3/2}$ & 6.40 & $\cdots$ &       &     &    \\
     \, 90 &  37.94 &  37.92 & \ion{Mg}{    11} ? &                                        $\;1s\;4p^{}\;^{3}P_{2}\; \rightarrow \;1s\;2s^{}\;^{3}S_{1}$ & 6.80 &       56 & $\pm$ &  11 &    \\
     \, 91 &  40.31 &  40.27 & \ion{C }{     5} &                                           $\;1s\;2p^{}\;^{1}P_{1}\; \rightarrow \;1s^{2}\;^{1}S_{0}$ & 6.00 &       38 & $\pm$ &  10 &  Y \\
     \, 92 &  43.76 &  43.76 & \ion{Si}{    11} &                                           $\;2s\;3p^{}\;^{1}P_{1}\; \rightarrow \;2s^{2}\;^{1}S_{0}$ & 6.20 &       55 & $\pm$ &  11 &  Y \\
     \, 93 &  44.01 &  44.02 & \ion{Si}{    12} &                            $\;1s^{2}\;3d^{}\;^{2}D_{3/2}\; \rightarrow \;1s^{2}\;2p^{}\;^{2}P_{1/2}$ & 6.30 &       52 & $\pm$ &  11 &  Y \\
     \, 94 &  44.16 &  44.17 & \ion{Si}{    12} &                            $\;1s^{2}\;3d^{}\;^{2}D_{5/2}\; \rightarrow \;1s^{2}\;2p^{}\;^{2}P_{3/2}$ & 6.30 &      115 & $\pm$ &  14 &  Y \\
     \, 95 &  45.50 &  45.52 & \ion{Si}{    12} &                            $\;1s^{2}\;3s^{}\;^{2}S_{1/2}\; \rightarrow \;1s^{2}\;2p^{}\;^{2}P_{1/2}$ & 6.30 &       49 & $\pm$ &  11 &    \\
     \, 96 &  45.68 &  45.69 & \ion{Si}{    12} &                            $\;1s^{2}\;3s^{}\;^{2}S_{1/2}\; \rightarrow \;1s^{2}\;2p^{}\;^{2}P_{3/2}$ & 6.30 &       57 & $\pm$ &  11 &  Y \\
     \, 97 &  46.30 &  46.30 & \ion{Si}{    11} &                                        $\;2s\;3d^{}\;^{3}D_{2}\; \rightarrow \;2s\;2p^{}\;^{3}P_{1}$ & 6.20 &       31 & $\pm$ &  10 &    \\
     \, 98 &  46.42 &  46.40 & \ion{Si}{    11} &                                        $\;2s\;3d^{}\;^{3}D_{3}\; \rightarrow \;2s\;2p^{}\;^{3}P_{2}$ & 6.20 &       41 & $\pm$ &  10 &  Y \\
     \, 99 &  49.17 &  49.18 & \ion{Si}{    11} &                                        $\;2s\;3s^{}\;^{3}S_{1}\; \rightarrow \;2s\;2p^{}\;^{3}P_{2}$ & 6.20 &       28 & $\pm$ &  10 &    \\
       100 &  49.20 &  49.22 & \ion{Si}{    11} &                                        $\;2s\;3d^{}\;^{1}D_{2}\; \rightarrow \;2s\;2p^{}\;^{1}P_{1}$ & 6.20 &       35 & $\pm$ &  11 &    \\
       101 &  50.33 &  & \multicolumn{2}{l}{\ion{Fe}{17} $\lambda$ 16.78 3rd order} &  &       52 & $\pm$ &  11 &    \\
       102 &  56.90 &  & \multicolumn{2}{l}{\ion{O}{8} Ly$\alpha$ 3rd order} & &       65 & $\pm$ &  11 &    \\
       103 &  57.71 &  &          No Id &                                                                                                  ... & &       23 & $\pm$ &   9 &    \\
       104 &  66.30 &  66.25 & \ion{Fe}{    16} &                                            $\;4f^{}\;^{2}F_{5/2}\; \rightarrow \;3d^{}\;^{2}D_{3/2}$ & 6.50 &       23 & $\pm$ &   9 &  \\
\hline
\end{tabular}
\normalsize
\end{table*}
\begin{table*}[!h]
\centering
\parbox{14cm}{
\bf{Table~\ref{tab:lines}.}\small\rm~Continued
\vspace{5mm}
}
\small
\begin{tabular}[h]{@{\hspace{4mm}}l@{\hspace{4mm}}rrllcr@{\hspace{1mm}}c@{\hspace{1mm}}lc}
\hline
\hspace{-2mm}$Label^{a}$ & $\lambda_{\mathrm{obs}}^{b}$\hspace{-3mm} & $\lambda_{\mathrm{pred}}^{b}$ & \multicolumn{1}{c}{Ion} & \multicolumn{1}{l}{Transition $(upper \rightarrow lower)$} & $\log T_{\mathrm{max}}^{c}$ & $(F\,$ & $\pm$ & $\sigma)^{d}$ & $EM^{e}$ \\
\hline
       105 &  66.37 &  66.36 & \ion{Fe}{    16} &                                            $\;4f^{}\;^{2}F_{7/2}\; \rightarrow \;3d^{}\;^{2}D_{5/2}$ & 6.50 &       36 & $\pm$ &  10 &  Y \\
       106 &  88.09 &  88.08 & \ion{Ne}{     8} &                            $\;1s^{2}\;3p^{}\;^{2}P_{3/2}\; \rightarrow \;1s^{2}\;2s^{}\;^{2}S_{1/2}$ & 5.80 &       37 & $\pm$ &  10 &  Y \\
       107 &  88.14 &  88.12 & \ion{Ne}{     8} &                            $\;1s^{2}\;3p^{}\;^{2}P_{1/2}\; \rightarrow \;1s^{2}\;2s^{}\;^{2}S_{1/2}$ & 5.80 &       40 & $\pm$ &  10 &  Y \\
       108 &  93.96 &  93.92 & \ion{Fe}{    18} &                              $\;2s\;2p^{6}\;^{2}S_{1/2}\; \rightarrow \;2s^{2}\;2p^{5}\;^{2}P_{3/2}$ & 6.80 &       60 & $\pm$ &  11 &  \\
      109a &  98.29 &  98.26 & \ion{Ne}{     8} &                            $\;1s^{2}\;3d^{}\;^{2}D_{5/2}\; \rightarrow \;1s^{2}\;2p^{}\;^{2}P_{3/2}$ & 5.80 &       34 & $\pm$ &   9 &    \\
      109b & $\cdots$ &  98.27 & \ion{Ne}{     8} &                            $\;1s^{2}\;3d^{}\;^{2}D_{3/2}\; \rightarrow \;1s^{2}\;2p^{}\;^{2}P_{3/2}$ & 5.80 & $\cdots$ &       &     &    \\
       110 & 108.41 & 108.36 & \ion{Fe}{    19} &                                  $\;2s\;2p^{5}\;^{3}P_{2}\; \rightarrow \;2s^{2}\;2p^{4}\;^{3}P_{2}$ & 6.90 &       48 & $\pm$ &  10 &    \\
       111 & 109.99 & 109.95 & \ion{Fe}{    19} &                                  $\;2s\;2p^{5}\;^{3}P_{1}\; \rightarrow \;2s^{2}\;2p^{4}\;^{3}P_{0}$ & 6.90 &       10 & $\pm$ &   7 &    \\
      112a & 110.65 & 110.63 & \ion{Fe}{    20} &                              $\;2s\;2p^{4}\;^{2}D_{3/2}\; \rightarrow \;2s^{2}\;2p^{3}\;^{2}D_{3/2}$ & 7.00 &        9 & $\pm$ &   7 &    \\
      112b & $\cdots$ & 110.63 & \ion{Ne}{     7} &                                           $\;2p\;3d^{}\;^{3}D_{3}\; \rightarrow \;2p^{2}\;^{3}P_{2}$ & 5.80 & $\cdots$ &       &     &    \\
       113 & 114.47 & 114.41 & \ion{Fe}{    22} &                               $\;2s\;2p^{2}\;^{2}P_{3/2}\; \rightarrow \;2s^{2}\;2p^{}\;^{2}P_{3/2}$ & 7.10 &       10 & $\pm$ &   7 &  \\
       114 & 117.16 & 117.18 & \ion{Fe}{    22} &                               $\;2s\;2p^{2}\;^{2}P_{1/2}\; \rightarrow \;2s^{2}\;2p^{}\;^{2}P_{1/2}$ & 7.10 &       33 & $\pm$ &   9 &  \\
       115 & 121.21 & 121.20 & \ion{Fe}{    21} &                                  $\;2s\;2p^{3}\;^{3}P_{2}\; \rightarrow \;2s^{2}\;2p^{2}\;^{3}P_{2}$ & 7.00 &       12 & $\pm$ &   7 &    \\
       116 & 121.85 & 121.84 & \ion{Fe}{    20} &                              $\;2s\;2p^{4}\;^{4}P_{3/2}\; \rightarrow \;2s^{2}\;2p^{3}\;^{4}S_{3/2}$ & 7.00 &       19 & $\pm$ &   8 &    \\
       117 & 128.81 & 128.75 & \ion{Fe}{    21} &                                  $\;2s\;2p^{3}\;^{3}D_{1}\; \rightarrow \;2s^{2}\;2p^{2}\;^{3}P_{0}$ & 7.00 &       35 & $\pm$ &   9 &  Y \\
      118 & 132.92 & 132.84 & \ion{Fe}{    20} &                              $\;2s\;2p^{4}\;^{4}P_{5/2}\; \rightarrow \;2s^{2}\;2p^{3}\;^{4}S_{3/2}$ & 7.00 &       24 & $\pm$ &   9 &    \\
%      118b & $\cdots$ & 132.91 & \ion{Fe}{    23} &                                           $\;2s\;2p^{}\;^{1}P_{1}\; \rightarrow \;2s^{2}\;^{1}S_{0}$ & 7.10 & $\cdots$ &       &     &    \\
       119 & 132.96 & 132.91 & \ion{Fe}{    23} &                                           $\;2s\;2p^{}\;^{1}P_{1}\; \rightarrow \;2s^{2}\;^{1}S_{0}$ & 7.10 &       39 & $\pm$ &  10 &  Y \\
       120 & 135.88 & 135.76 & \ion{Fe}{    22} &                               $\;2s\;2p^{2}\;^{2}D_{3/2}\; \rightarrow \;2s^{2}\;2p^{}\;^{2}P_{1/2}$ & 7.10 &       17 & $\pm$ &   8 &  \\
\hline
\end{tabular}
\renewcommand{\baselinestretch}{1.2}
\parbox{14cm}{
\vspace{2mm}
\small
$^{a}$ Invididual components of unresolved line blends are indicated by letters.

$^{b}$ Observed and predicted (CHIANTI database) wavelengths.

$^{c}$ Maximum emissivity temperature in K.

$^{d}$ Total line counts and errors.

$^{e}$ Lines selected to derive the emission measure distribution.

}
\normalsize
\end{table*}

\clearpage

contaminating iron line because all of the other prominent
\ion{Fe}{17} lines in the spectrum are well predicted (within a factor 1.5).
Other possible explanations include optical thickness effects 
due to resonant line scattering \citep{tdpd04} or
uncertainties in the theoretical line ratios due to processes of
population of the excited atomic levels not taken into account by the
adopted line emissivity code \citep{sbl+01}.

\begin{figure}[!ht]
   \plotone{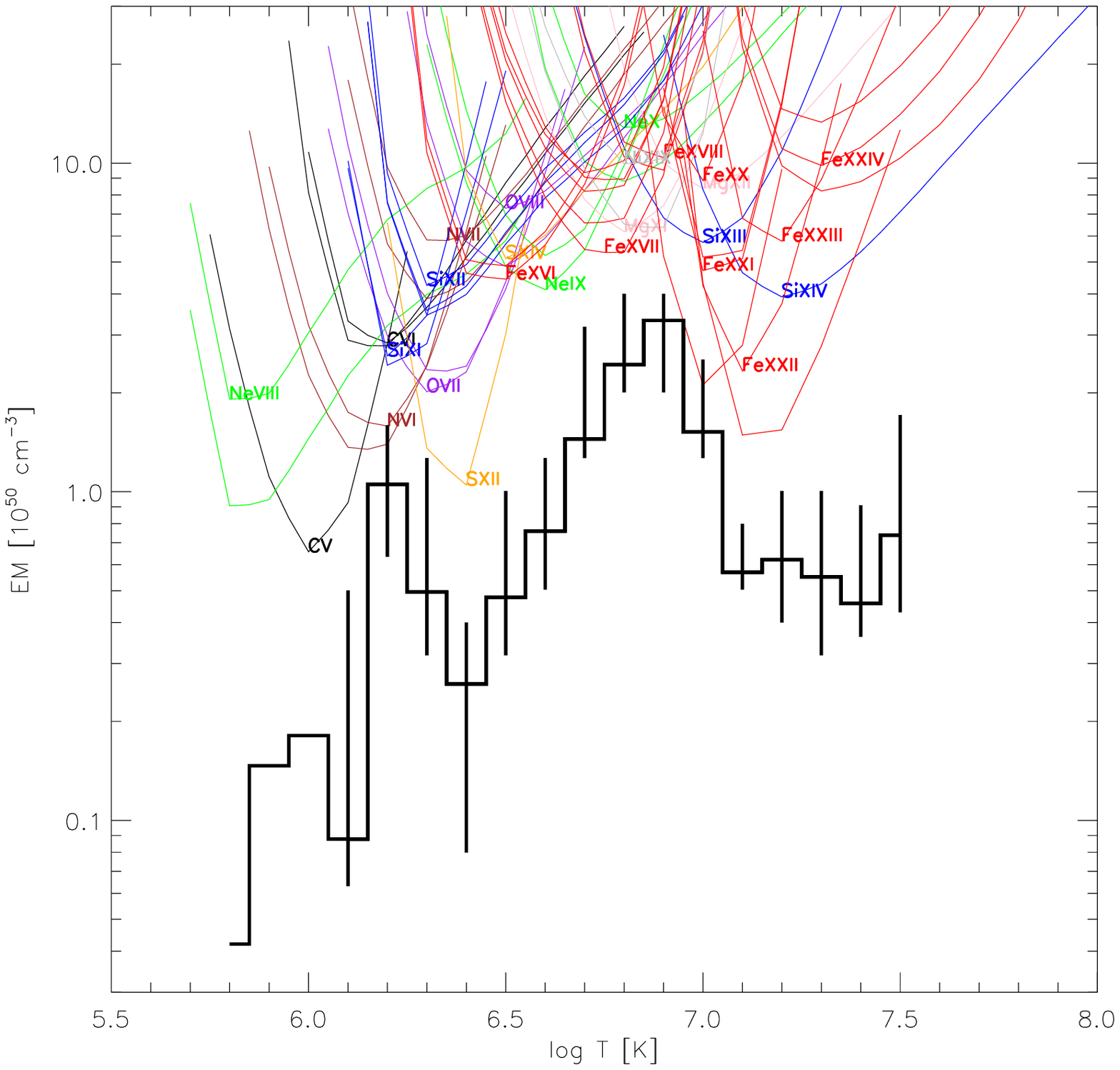}
%  \centering
%  \resizebox{\hsize}{!}{
%  \includegraphics[bb=54 85 558 550]{f3.eps}
%  }
\caption{
Emission measure distribution (solid histogram) vs.\ temperature
and inverse line emissivity curves,
scaled by the individual element abundances relative to iron.
Representative ion labels are placed at the relevant peak emissivity
temperature. In the electronic edition, the colors of the emissivity curves
indicate different elements (C black, N brown, O purple, Ne green, Mg
pink, Si blue, S orange, Fe red, Ni gray).
\label{fig:em}}
\end{figure}

The final EMD is shown in
Fig.~\ref{fig:em}, together with the inverse emissivity curves computed with
the set of abundances (relative to Fe) reported in Table~\ref{tab:ab}. 
Note that the
available data allow us to constrain the emission measure distribution in the
temperature range $\log T \sim$ 6.0--7.5\,K.
Note also that if the \ion{Ne}{10} Ly$\alpha$ doublet is used for the
emission measure analysis, the resulting EMD in the temperature range
$\log T$~6.4--6.9\,K turns out to be flatter
than the one in Fig.~\ref{fig:em}, but compatible within formal
uncertainties, and the Ne/Fe abundance ratio lower by a factor 1.6
($Ne/Fe \sim 3$ using the \ion{Ne}{10} Ly$\alpha$, 
rather than $Ne/Fe \lesssim 5$).

\begin{figure}[!ht]
\plotone{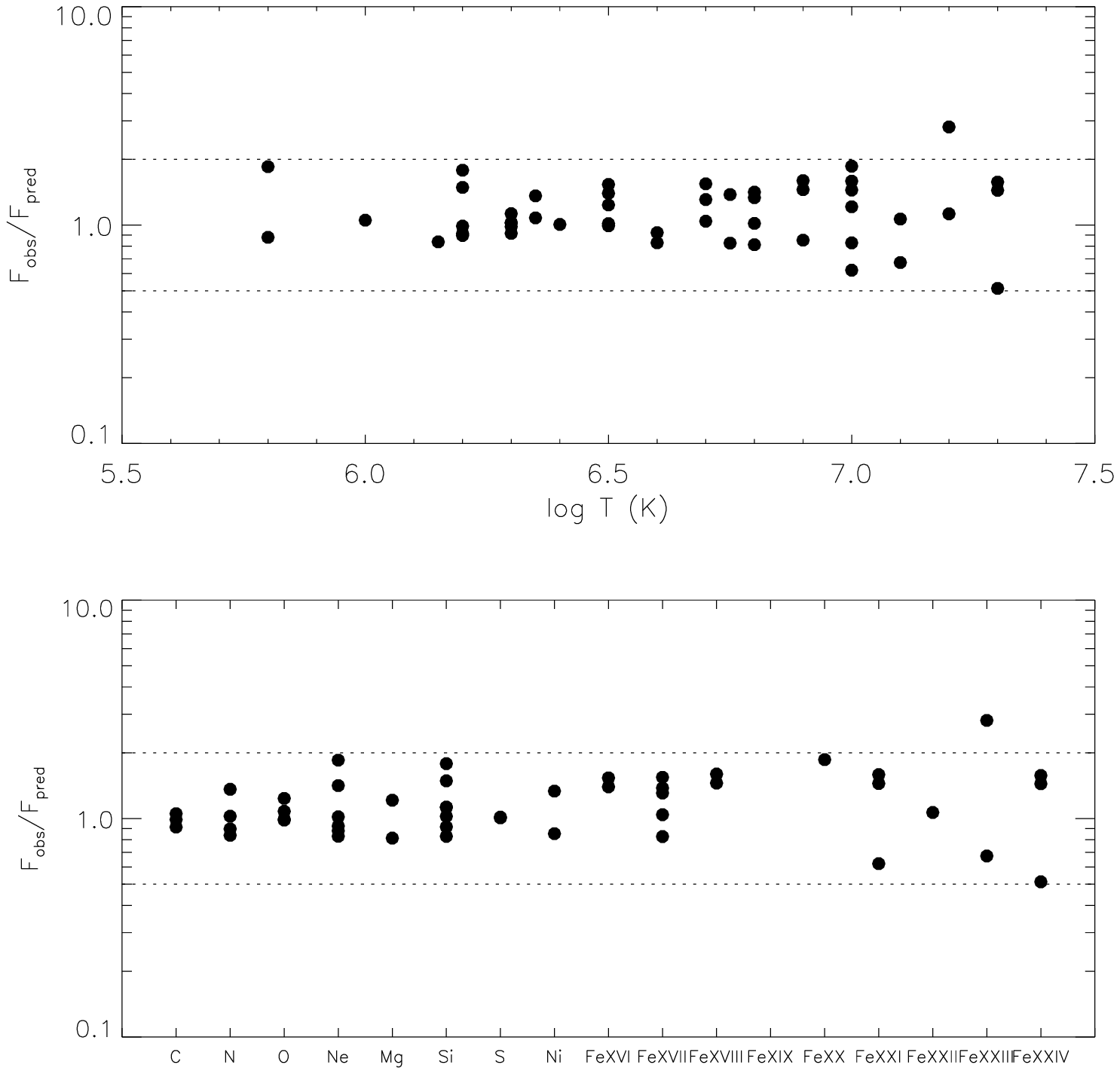}
\caption{
Ratios of observed to predicted line fluxes vs.\ temperature (upper panel) and 
vs.\ ionization stage (lower panel). 
The symbols refer to lines with relatively high
S/N ratio selected for the emission measure analysis; the
agreement between measurements and model predictions is generally 
within a factor 2 (dotted horizontal lines).
% Open symbols indicate other measured lines with more than 20 integrated
% counts (circles) or between 10 and 20 integrated counts (triangles).
\label{fig:flx}}
\end{figure}

The appropriateness of our solution can be best appreciated from inspection of
Fig.~\ref{fig:flx} where the ratio of observed to predicted line counts vs.\
temperature and vs.\ ionization species is shown: 
the agreement is within a factor 2 for all of the
selected lines, with the exception of the \ion{Fe}{22} line at
$\lambda$ 11.75\,\AA. A more general qualitative comparison of the observed
and predicted model spectrum (folded with the analytical line spread
function used for line fitting) shows a good overall agreement, but
also some deficiency of the {\sc chianti} spectrum in several spectral
regions populated by many, relatively weak emission lines from L-shell
transitions of \ion{Fe}{17}--\ion{}{19} ions: in particular,
we note the 10--12\,\AA\ range -- where the most prominent \ion{Fe}{24}
lines are located, but where several contaminating lines from lower
ionization species are missing from the {\sc chianti} database -- and the
region of the \ion{Ne}{9} triplet (13.4--13.8\,\AA), where similar
incompleteness effects are visible. By employing the Astrophysical
Plasma Emission Database ({\sc aped}) V1.3
line list and emissivities, we have verified
that our EMD and chemical abundances yield a
predicted spectrum that matches the available data better in both
wavelength regions mentioned above. The robustness of our
solution derives from the careful selection of strong and reliable lines
for the emission measure analysis. 
 
\begin{figure}[!ht]
   \plotone{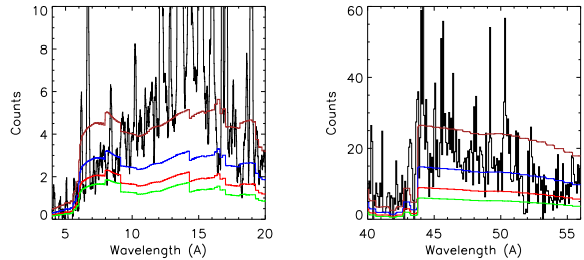}
%  \centering
%  \resizebox{\hsize}{!}{
%  \includegraphics[bb=56 400 509 585]{f5.eps}
%  }
\caption{
Comparison between the observed spectrum (with a bin size
of 0.075\,\AA) and the continuum emission
level predicted for different values of the 
Fe/H abundance, with the four stacked curves
corresponding from top to bottom to: 0.2, 0.4, 0.8,
and 1.6 times the solar metallicity (brown, blue, red, and green,
respectively, in the electronic edition).
\label{fig:z}}
\end{figure}

In order to determine the $Fe/H$ ratio in the corona of AD~Leo,
we have performed a detailed comparison (Fig.~\ref{fig:z}) 
between the observed continuum emission
level and that predicted with different assumptions of the plasma
metallicity, $Z$. In fact, the relative line to continuum ratio increases with
increasing $Z$, while the volume emission measure needs to be scaled
down by the same factor in order to keep the total predicted source
flux constant. A close inspection of the spectral regions 6--9\,\AA,
18--21\,\AA, and 44--54\,\AA, where the continuum is more visible, 
suggests a metallicity larger than $0.8 Z_{\sun}$
for the corona of AD~Leo; as explained in Sect.~\ref{sec:metal},
this relatively high value implies that the $Fe/H$ ratio
cannot be determined to within less than a factor 2 (Table~\ref{tab:ab}).

The {\it Chandra} LETG spectrum of AD~Leo allowed us to estimate the plasma
density at different temperatures from sensitive line ratios derived 
from the \ion{O}{7}, \ion{Ne}{9}, and \ion{Si}{13} He-like triplets,
and also from \ion{Fe}{20}, \ion{Fe}{21}, and \ion{Fe}{22} lines.
The details of the analysis are described in Sect.~\ref{sec:dens},
and the main results are reported in Table~\ref{tab:tri}.

\clearpage

\begin{deluxetable}{lrrll}
\tablewidth{0pt}
\tablecaption{Coronal abundances in AD Leo
\label{tab:ab}}
\tablecolumns{5}
\tablehead{
 & \multicolumn{1}{c}{FIP} \\
 & \multicolumn{1}{c}{eV} & 
$N$\tablenotemark{a} &
\multicolumn{1}{c}{$A_{\rm Z}/A_{\rm H}$\tablenotemark{b}} & 
\multicolumn{1}{c}{$A_{\rm Z}/A_{\rm Z\odot}$\tablenotemark{c}}
}
\startdata
C  & 11.26 & 3       & $(\,9.6^{+3.3}_{-1.4}\,)\times10^{-4}$     & $2.7^{+0.9}_{-0.4}$ \\
N  & 14.53 & 4       & $(\,3.0^{+1.0}_{-0.5}\,)\times10^{-4}$     & $3.3^{+1.1}_{-0.5}$ \\
O  & 13.62 & 4       & $(\,1.6^{+0.5}_{-0.3}\,)\times10^{-3}$     & $2.1^{+0.7}_{-0.4}$ \\
Ne & 21.56 & 6       & $(\,5.6^{+0.9}_{-1.5}\,)\times10^{-4}$     & $4.7^{+0.7}_{-1.3}$ \\
Mg &  7.65 & 2       & $(\,3.0^{+2.1}_{-0.5}\,)\times10^{-5}$     & $0.8^{+0.6}_{-0.1}$ \\
Si &  8.15 & 7       & $(\,9.0^{+1.3}_{-2.5}\,)\times10^{-5}$     & $2.5^{+0.4}_{-0.7}$ \\
S  & 10.36 & 2       & $(\,1.1^{+1.0}_{-1.4}\,)\times10^{-4}$     & $6.6^{+6.0}_{-1.6}$ \\
\medskip
Ni & 7.64 & 2 	     & $(\,3.4^{+0.7}_{-1.4}\,)\times10^{-6}$     & $1.9^{+0.4}_{-0.8}$ \\
Fe\tablenotemark{d} & 7.90 & 18 & \multicolumn{1}{l}{\phantom{(\,}2.6 -- 5.2$\,\times10^{-5}$} & \multicolumn{1}{l}{\hspace{-2mm}0.8 -- 1.6} \\
\enddata
\tablenotetext{a}{Number of spectral lines of the relevant element included
in the emission measure reconstruction.}
\tablenotetext{b}{Element abundances relative to hydrogen.}
\tablenotetext{c}{Element abundances relative to the solar ones \citep{gns+92}.}
\tablenotetext{d}{Abundance range determined by matching the continuum
(Sect.~\ref{sec:metal}).}
\end{deluxetable}

\begin{deluxetable}{lcrcccc}
\tablecaption{Density and temperature spectroscopic diagnostics\tablenotemark{a}
\label{tab:tri}}
\tablehead{
\colhead{Ion} & 
\colhead{$R_{D}$\tablenotemark{b}} &
\colhead{$N_{\mathrm{e}}$}        &
\colhead{$R_{T}$\tablenotemark{c}} &
\colhead{$\log T$} &
\colhead{$\log T_{\mathrm{max}}$\tablenotemark{d}} &
\colhead{$P$\tablenotemark{e}}\\
\colhead{} &
\colhead{} &
\colhead{cm$^{-3}$} &
\colhead{} &
\colhead{K} &
\colhead{K} &
\colhead{dyn cm$^{-2}$}}

\startdata
\ion{O}{7}  & $3.15\pm0.68$ & $(7.1^{+8.2}_{-6.0})\times10^{9\phn}$ &
              $0.86\pm0.12$ & $6.29\pm0.12$ & 6.3 & $3.8^{+7.0}_{-3.4}$ \\
\ion{Ne}{9} & $1.88\pm0.37$ & $(3.9^{+3.7}_{-1.6})\times10^{11}$ &
              $1.02\pm0.15$ & $< 6.44$      & 6.6 & $< 3 \times 10^2$ \\
\ion{Si}{13}& $0.92\pm0.38$ & $(9.8^{+9.2}_{-6.1})\times10^{13}$ &
              $1.03\pm0.35$ & $< 6.90$      & 7.0 & $< 4 \times 10^5$ \\
\ion{Fe}{21}& $0.35\pm0.22$ & $(1.8^{+2.3}_{-1.5})\times10^{12}$ &
              \nodata       & $7.02$\tablenotemark{f} & 7.0 &
$5^{+7}_{-4} \times 10^3$ \\
\enddata
\tablenotetext{a}{Statistical uncertainties are all at the
$1\sigma$ confidence level} 
\tablenotetext{b}{Density sensitive f/i line ratios}
\tablenotetext{c}{Temperature sensitive (f+i)/r line ratios}
\tablenotetext{d}{Peak emissivity temperature}
\tablenotetext{e}{Plasma pressure, $P = 2 N_{\mathrm{e}} k_{\mathrm{B}} T$}
\tablenotetext{f}{Average temperature, weighted by the line emissivity
and the emission measure distribution vs. temperature}
\end{deluxetable}

\clearpage

\section{Discussion}
\label{sec:discuss}

\subsection{Plasma thermal structure and abundances}

\begin{figure}[!ht]
\plotone{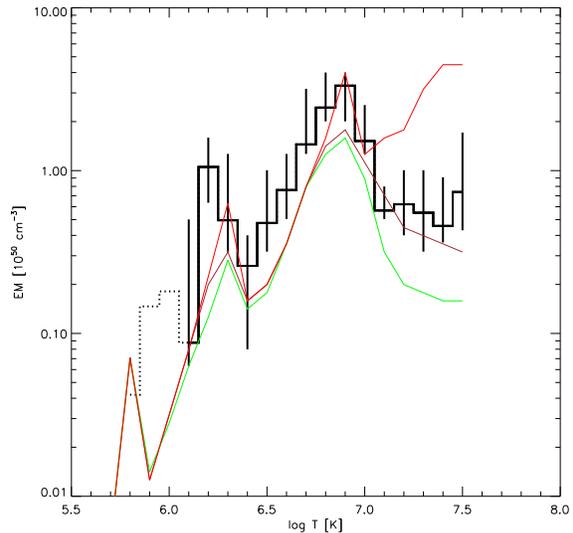}
\caption{
Comparison between the {\it Chandra}
emission measure distribution (thick-line histogram) and the three
distributions derived by \protect\citet{sm02} during quiescent, average, and
flaring phases (from bottom to top, and colored green, brown, and red in
the electronic edition), based on EUVE observations taken
between 1993 and 2000.
\label{fig:jsanz}}
\end{figure}

Our reconstructed EMD
indicates that the corona of AD Leo is dominated by plasma in the temperature 
range $3 \times 10^6$ -- $10^7$\,K, in qualitative agreement with previous 
results based on EUVE and SAX data \citep{cfh+97,smf99},
but a sizeable fraction ($\sim 20$\%) of the plasma volume emission measure
is at higher temperatures \citep[up to $\sim 30$\,MK; see also][]{brm+03}.
Figure~\ref{fig:jsanz} shows a comparison with the three EMDs
derived by \citet{sm02} during quiescent and low-amplitude flaring 
phases of AD~Leo, and in its average state (based on the sum of all the
nine EUVE spectra discussed by these authors). 
Our {\it Chandra} EMD turns out to be remarkably similar to the latter,
but higher by a factor $\lesssim 2$ in the temperature range 3--6\,MK.
The overall agreement between these EMDs, based on data of different
instruments and derived with different procedures, strengthens 
our confidence in the results of the line-based emission measure analysis.
This also suggests that
the corona of AD~Leo has maintained nearly
the same configuration on time scales of about 10 years.

The EMD of AD~Leo appears similar to that of other active 
late-type stars but is significantly less steep.
The volume emission measure increases as $T^{\beta}$ in the range 
$\log T = 6.4$--6.9, with $\beta = 2.2 \pm 1.1$, as
compared with slopes $\beta = 4$--6 observed in RS CVn-type binaries
\citep*{gj98,sbd03,amp03}, and also in single late-type dwarfs and 
yellow giants \citep{ass+98,dpo+00,smp+04}. This result implies that
the bulk of the X-ray emission of AD~Leo can be explained by
plasma confined in static, constant cross-section and uniformly-heated
coronal 
loop structures (see also below).
Before attempting an interpretation with such a model, however, we must
address the issue of the coronal abundances.

\begin{figure}[!ht]
\plotone{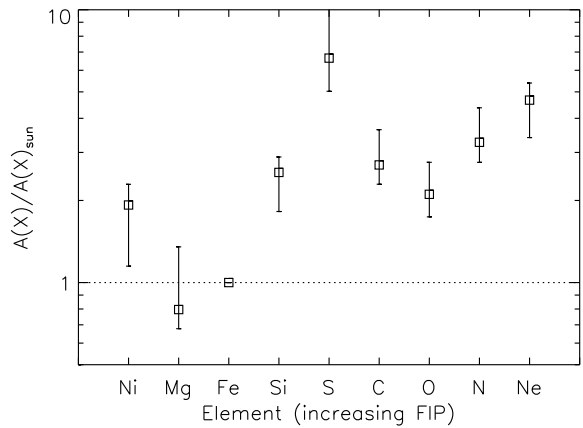}
\caption{
Pattern of coronal abundances (relative to solar values) vs. chemical
element sorted by First
Ionization Potential, assuming solar Fe abundance
(for different Fe values, all points shift by the same factor).
\label{fig:ab}}
\end{figure}

Abundances in the corona of AD~Leo (Table~\ref{tab:ab})
are all larger than the solar photospheric 
values of \citet{gns+92} by factors of 2--4, with
the exception of Mg and Fe, whose abundances are solar within errors, 
and of S, which appears overabundant compared to the solar mixture.
These chemical abundances when sorted by the First Ionization
Potential (FIP) of each element reveal the characteristic
pattern shown in Fig.~\ref{fig:ab}.
In the solar corona, and in particular in long-lived coronal structures,
the composition of the plasma appears enriched by
low-FIP elements (FIP $< 10$\,eV) by about 
a factor 4 (on average) with respect to photospheric values \citep{fl00}, 
while in other stars a more complex behavior has been observed 
\citep*{d03,sfm04}, with a tendency for the low-FIP elements (including
iron) to become depleted with respect to the high-FIP elements (Neon
in particular) in extremely active RS CVn-type and Algol-type
binaries.  At face value, in the corona of AD~Leo there is some
evidence for a FIP-related bias in the abundances, similar to that of
other stars with high activity levels, although abundances relative to
stellar {\em photospheric} values should be employed to confirm such a
trend.  

Unfortunately, stellar photospheric spectra of dM stars are
notoriously difficult to analyze, and firm measurements of
photospheric abundances in AD~Leo are limited to studies by
\citet*{nsp92} and \citet{jlah96}, who estimate $[Fe/H] = -0.47$ and
$[Fe/H] = -0.75 \pm 0.25$, respectively.  These values can probably be
revised upward by 0.17~dex compared to our abundances: in the case of
\citet{jlah96} it is not clear which solar Fe abundance was assumed,
though \citet{nsp92} adopted $\log A_{\rm Fe} = 7.67$ (in the conventional
spectroscopic logarithmic units), compared to $\log A_{\rm Fe} = 7.51$ 
recommended by \citet{gns+92}. 
Both \citet{nsp92} and \citet{jlah96} remark on the
apparent metal paucity of AD~Leo for its apparent age: 
in fact, AD~Leo is a population I object by its motion, 
and quite young by its fast rotation and enhanced activity.  
Our coronal abundances suggest, then, that either the corona is generally 
enhanced in metals relative to photosphere, or else the existing photospheric
estimates are spuriously low.  The latter is our preferred explanation
owing to the difficulties inherent in the compositional analysis of such 
an active dMe star.

The most puzzling result, however, is the relatively high metallicity
(represented by a nearly solar Fe/H value) suggested by the
line-to-continuum ratios (Sect.~\ref{sec:sp} and Fig.~\ref{fig:z}) as
compared to previous low resolution X-ray studies: although the result
is affected by a large uncertainty, Fe/H values 0.2--0.3 times solar
\citep{smf99,fmr00} seem to be incompatible with our data.
\citet{d02} noted that there can be a tendency of finding spuriously
low abundances with global model fitting approaches as applied to
lower resolution spectra: both missing line transitions in spectral
models that form a pseudo-continuum and limited signal-to-noise ratio
can result in line flux being interpreted as continuum.  However, the
size of the discrepancy between our abundance results and those
derived from low resolution spectra is still surprising.  We note that
this uncertainty affects the absolute scale of the other element
abundances but not their ratios, and also the absolute value of the
plasma emission measure but not the shape of its distribution vs.\
temperature.

Of the individual abundance results, the high S and Si abundances are
notable.  With FIP$=10.36$\,eV, S is generally considered an
intermediate FIP element, and is generally not seen to be
conspicuously over- or underabundant.  If the relative abundances of
elements in the photosphere of AD~Leo represent those of the solar
mixture, as the analysis of \citet{nsp92} suggests, then AD~Leo would
be the first active star for which a relative S enhancement larger
than that of Ne has been found.

In the case of Si, an enhancement relative to Fe and Mg is
particularly interesting because these elements all have very similar
FIP. \citet{d02} also found evidence for variation in the coronal
Si/Mg ratio in other active stars, suggesting it might be a common
coronal phenomenon.  
%While again we have to caution against premature
%interpretation of coronal abundances in terms of anomalies for stars
%in which photospheric abundances have not been accurately determined
An enhancement of Si relative to Fe might
possibly be understood in terms of gravitational settling, owing to
the different masses of these elements \citep[e.g.][]{d03}; however, Mg
and Si also have very similar masses and any fractionation mechanism
based on charge, mass or charge/mass ratio would need to be very
highly tuned.

\subsection{Physical interpretation}
\label{sec:phys}

A number of key studies based on previous observations at X-ray and EUV
wavelengths provide us with different working hypotheses on the corona
of AD Leo, and our aim is to test these against our new results.
\citet{grk+96} performed fits of
low-resolution ROSAT PSPC spectra of AD Leo and other M-type dwarfs,
together with variability analyses of their
X-ray light curves, and proposed possible modeling of 
the results in terms of two classes of coronal structures 
with different plasma temperatures and surface filling factors:
very small ($L \ll R_*$) magnetic loops in low-temperature ($T
\sim 2 \times 10^6$\,K) quiescent active regions,
and possibly small ($L \lesssim R_*$), unstable (flaring), 
high-temperature ($T \sim 10^7$\,K) loops. A few years later,
a detailed coronal loop model analysis of BeppoSAX spectra was
presented by \citet{smf99}, who inferred that the dominant class of
X-ray emitting structures was composed of loops with maximum plasma
temperature $T_{\rm max} \sim 10^7$\,K with sizes $L < 0.1 R_*$,
covering no more than 1\% of the stellar surface. Those data also suggested a second class of
higher temperature structures that were not
as well constrained as in the former class.  
A detailed analysis of several X-ray flares allowed \citet{fmr00} to 
measure peak temperatures of 10-50 MK and
to infer sizes of the flaring loop structures spanning the range
4--13$\times 10^9$\,cm (i.e., 0.15--0.5\,$R_*$). By applying the
\citet{rtv78} scaling laws for isobaric loops with constant
cross-section and uniform heating, \citet{fmr00} also derived
plasma pressures in the range $\approx 10^2$--$10^4$\,dyn cm$^{-2}$. 

The flaring loops pinpointed by \citet{fmr00} are compatible with the
high-temperature coronal structures suggested by \citet{grk+96}, but 
they are not as compact as indicated by the loop modeling of \citet{smf99}. 
On the other hand, it is not yet clear whether the same class of flaring loops, 
in their quiescent state, may explain the ``stable'' part of the EMD of
AD~Leo, or whether they only populate its high-temperature tail.
Finally, the very small and cool loops found by \citet{grk+96} were not
probed by the other two studies, possibly due to selection effects, 
namely the different bandpasses of the instruments and their relative 
sensitivity to cool plasma emission.

\begin{figure}[!ht]
\plotone{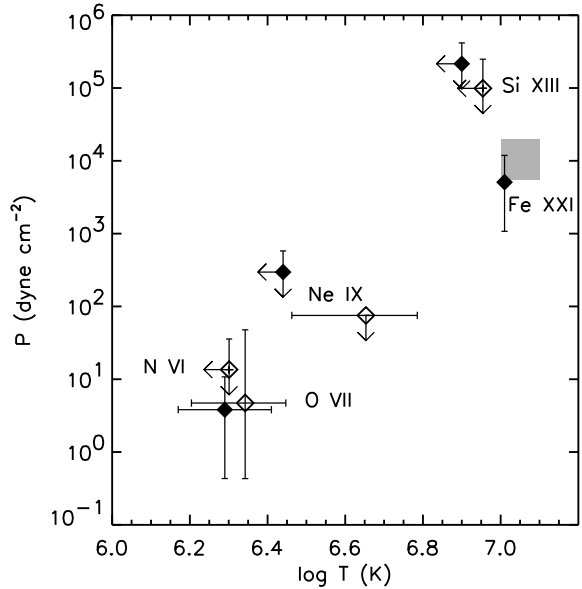}
\caption{
Plasma pressures (or upper limits) vs.\ temperature 
(with associated $1\sigma$ error bars)
derived from density-sensitive spectroscopic diagnostics 
(He-like triplets and \ion{Fe}{21} lines). Error bars
are based on $1\sigma$ statistical uncertainties on line fluxes.
Open symbols refer to measurements presented by \protect\citet{nsb+02};
shaded area shows measurements based on \ion{Fe}{20}--\ion{}{22} line
ratios derived by \protect\citet{sm02} from EUVE spectra.
\label{fig:press}}
\end{figure}

A crucial new piece of information is the plasma pressure, 
which we have computed from the measured densities and temperatures 
(Table~\ref{tab:tri}) consistently derived from spectroscopic diagnostics. 
A plot of the pressure vs.\ temperature is shown in Fig.~\ref{fig:press},
which includes values computed from the results of
\citet{nsb+02} (based on the same {\it Chandra}/LETG observation), and the
range of pressures spanned by the results of \citet{sm02} -- who employed
density-sensitive line ratios from \ion{Fe}{20}--\ion{}{22}
ions measured in EUVE spectra. The various results mentioned above are
in general agreement,
within $1\sigma$ statistical uncertainties,
with the exception of problematic points
related to the \ion{Ne}{9} triplet. In particular, the densities
based on ratios of \ion{Fe}{21} EUV lines consistently suggest plasma
pressures $\approx 10^4$\,dyn cm$^{-2}$ at $T \approx 10^7$\,K; in this
respect, the LETG spectrum of AD~Leo resolves the concern about possible
unaccounted blends of the iron lines in the lower-resolution EUVE data
and confirms the previous finding based on these data. 
At the low temperatures ($T \approx 2 \times 10^6$\,K) 
probed by the \ion{O}{7} triplet, instead, pressures of the order 1--10\pu 
are consistently found. 

The above result raises again the issue of the pressure balance in the
coronae of active stars, which we have already seen in previous works \citep{amp03,smm03}.
Unless the results based on spectroscopic diagnostics are affected by
systematic biases due to atomic physics uncertainties, the only viable
solution appears to be the existence of separate classes of coronal
loops with different plasma densities, contributing to the EMD
at different temperatures. As already suggested by
\citet{amp03} in the case of Capella, the high-pressure plasma could be
confined in a class of loop structures having maximum
temperatures near the peak of the EMD or
higher, while the low-pressure plasma could be identified with
structures determining the secondary EMD peak at $T \sim 2$\,MK.
While the latter appear similar to solar coronal active regions,
the former are only observed during the short time scales of solar flares
\citep*{por+00,rpo01}.

Since the EMD of AD~Leo appears very stable on time scales of several years,
we conclude that the X-ray emission of this active dMe star,
in its stationary state, is dominated by relatively hot and
high-pressure coronal structures, with respect to solar standards. 
Whether these structures are in a steady or dynamic state
is an issue which we address in the following section.
In particular, we will consider first the hypothesis that the corona
of AD~Leo, at any time, is dominated by flares and then the alternative
hypothesis that we are dealing with stationary (non-dynamic) coronal
structures.

\subsubsection{Continuous flaring hypothesis}

We start by noting that the flare peak temperatures reported by \cite{fmr00} 
fall in the high-temperature tail of the present EMD, and that the pressures 
are also within the range we have derived from the spectral diagnostics
(Tab.~\ref{tab:tri}).

One way of testing the ``continuous flaring'' hypothesis is by looking
for evidence of bulk plasma motions, as expected during the
chromospheric evaporation phase. To this aim we have looked for wavelength 
shifts of the emission lines observed in the range 80--140\AA, where the
spectral resolution of the LETG is highest. Some of
these lines appear redshifted by 0.02-0.05\AA, 
corresponding to downflow velocities of the order of $10^2$\,km s$^{-1}$, 
but residual uncertainties in the wavelength calibration and
the relatively low S/N ratio in this spectral region (see Fig.~\ref{fig:wvl}
in the Appendix~\ref{sec:id}) prevent us to draw any firm conclusion.

The second test we have considered is a comparison with model
predictions.  \citet{g+03} have recently investigated the flare rate
distribution of AD~Leo in EUV and X-ray radiated energy, 
from extensive observations with EUVE and BeppoSAX, and have derived a
power-law distribution, $dN/dE \propto E^{-\alpha}$ with $\alpha$ in the
range 2.0--2.5. This result implies that flares can play an important
role in the energy budget of the corona of AD~Leo. Following the
limiting hypothesis that the corona is entirely heated by flares, these
authors derived a time-averaged emission measure distribution
with a sharp rise in temperature and a less steep decline after the EMD
peak. The slopes of the ascending and descending segments depend on
several parameters of the model, and in particular the low-T part is
expected to rise as $T^{2/\zeta}$, with $\zeta$ being the slope of the
power-law dependence of the plasma temperature vs. density,
characteristic of the flare decay phase \citep*{rsp93}. 
By considering a
number of published values, \citet{g+03} propose $\zeta = 0.5$--1,
yielding $EMD \propto T^{3\pm1}$, which is consistent with the slope
$\beta$ we have derived above.

\subsubsection{Quasi-static corona hypothesis}

An alternative interpretation of the bulk of the EMD of AD~Leo is in
terms of quasi-static coronal loop structures, i.e., loops that evolve
on time scales longer than the characteristic cooling time scale of the
plasma \citep{srj+91}
\begin{equation}
\tau_{\rm c} \approx 120 \left( \frac{L}{10^9} \right)
\left( \frac{T}{10^7} \right)^{1/2} ~ {\rm sec.}
\label{eq:coolt}
\end{equation}
Since the pressure scale height at $T = 8$\,MK (EMD peak) is 
$H_{\rm p} \approx 1.6 \times 10^{10}$\,cm ($\sim 0.6 R_*$),
loops shorter than this size are in isobaric state; this was the case
for all the flaring structures studied by \citet{fmr00}. In this
hypothesis, the RTV scaling law
\begin{equation}
T_{\rm max} = 1.4\times 10^3 (p L)^{1/3}
\end{equation}
together with the gas relation $p = 2 N_{\rm e} k_{\rm B} T$, yields
\begin{equation}
N_{\rm e}^2 L = 1.7\times 10^{12} T_{\rm max}^4 L^{-1}
\end{equation}
which is approximately proportional to the emission measure 
per unit cross-sectional area of a single loop.
By modeling the corona as an ensemble of loops
all having the same characteristics and covering a fraction $f$ of the
visible  
hemisphere (hereafter, surface filling factor), we get a volume emission measure 
$EM = 2 \pi R_*^2 f N_{\rm e}^2 L$, which scales as
\begin{equation}
EM \simeq 1.1\times 10^{13} R_*^2 f T_{\rm max}^4 L^{-1}
\end{equation}
% where q is a geometry factor which accounts for the fraction of corona
% obscured by the stellar disk ($q = 0.5$ for $L \ll R_*$).
In a similar way, we can derive a scaling of the volume emission measure
with the maximum temperature and the pressure, which reads
\begin{equation}
EM \simeq 3.0\times 10^{22} R_*^2 f p T_{\rm max}
\label{eq:em_vs_pt}
\end{equation}
Hence, through knowledge of the EMD and the plasma
pressure (independently derived from spectroscopic diagnostics), we can
in principle estimate the surface filling factor of the X-ray emitting
coronal plasma. 

\begin{figure}[!ht]
\plotone{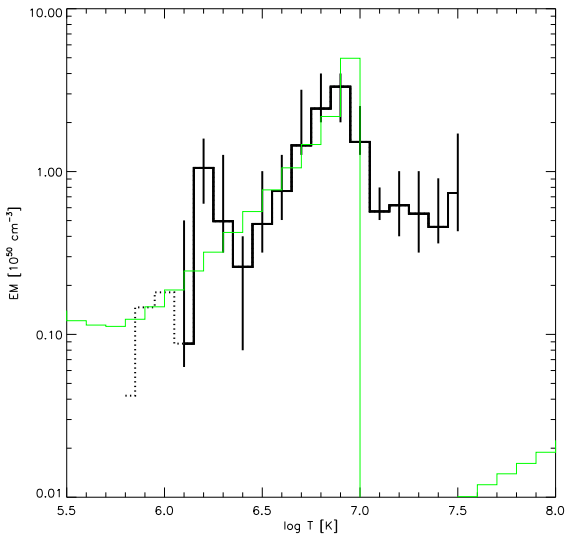}
\caption{
Comparison between the EMD derived from the {\it Chandra} data (heavy-line
histogram) and the 2-loop
model solution best-fitting the BeppoSAX spectra studied by
\protect\citet{smf99} (colored green in the electronic edition). 
The latter has been scaled to account for differences in
the plasma metallicity and in the stellar parameters (distance and radius).
\label{fig:loop}}
\end{figure}

The above coronal modeling, with the additional aspect of taking into account 
the detailed temperature and density profiles along the loops, was employed
by \citet{smf99} to synthesize the total X-ray emission of AD~Leo and to
fit the available BeppoSAX spectra\footnote{As mentioned above,
\citet{smf99} performed their spectral analysis with a 2-loop model,
i.e., consisting of two classes of coronal loops with different
characteristics. Here and in the following we refer only to the dominant
class, unless otherwise specified.}. No estimate of the loop length
or the plasma pressure was available at that time, and hence the best-fit
solution was not unique, in the sense that the
surface filling factor was constrained only by the requirement that the
loops were isobaric ($L < H_{\rm p}$). This limitation does not
affect the shape of the emission measure distribution but only its
absolute normalization, and hence we have attempted a direct comparison
of that coronal loop model solution with our current emission measure
distribution. Fig.~\ref{fig:loop} shows such a comparison, having
allowed for a down-scaling of the best-fit loop model solution by a factor 
$\sim 2$ to account for differences in the plasma metallicity and in some 
stellar parameters (distance and radius) with respect to the previous work.
The agreement is striking, and confirms that our previous
modeling was reasonable. 

We are now in the position to employ the
additional piece of information provided by our estimates of 
the plasma pressure. Since the emission measure distribution
peaks at $\approx 10^7$\,K, we can assume that the bulk of the coronal plasma 
is trapped in loops having this same value for $T_{\rm max}$, with a
plasma pressure comparable to that suggested by the density-sensitive
\ion{Fe}{21} lines, i.e., $p \approx 5 \times 10^3$\pu. We have computed a
loop model with the above parameters, yielding 
$L \sim 1.5 \times 10^8$\,cm, and the related EMD: a good match
with the reconstructed EMD of AD~Leo can be obtained with a surface
filling factor $f \sim 3 \times 10^{-4}$. This model suggests that, if
static loops are invoked to explain the X-ray emission of AD~Leo and
plasma pressures of the order of $10^3$\pu are assumed, very small 
($L \lesssim 10^{-2} R_*$) coronal structures are required, covering a
tiny fraction of the visible hemisphere ($f \ll 1$\%). For other parameter
values, equation (\ref{eq:em_vs_pt}) tells us that the surface filling factor
scales approximately as
\begin{equation}
f \approx 5 \times 10^{-4} \left( \frac{EM}{10^{50}} \right) \left(
\frac{p}{10^3} \right)^{-1} \left( \frac{T}{10^7} \right)^{-1}
\label{eq:f_vs_p}
\end{equation}

It is worth noting that the above model consists of structures much shorter
than the flaring loops studied by \cite{fmr00}, 
having $L \sim$ 0.15--0.5\,$R_*$. If we assume, alternatively, that the corona 
is composed of a population of loops with $L = 0.3 R_*$ and 
$T_{\rm max} = 10^7$\,K, we obtain a pressure $p \sim 45$\pu 
for the confined plasma and $f \sim 4$\%. This latter model appears
difficult to reconcile with the actual pressure values.

On the other hand, the low plasma pressure ($p \sim 4$\pu)
at $T \sim 2$\,MK, probed with the \ion{O}{7} triplet,
points toward the existence of an extensive population of cooler loops
which may explain the low-temperature tail of the EMD of AD~Leo.
If we postulate that the secondary EMD peak at $T \sim 1.6 \times 10^6$\,K
is at the maximum temperature of these cool loops, application
of equation (\ref{eq:f_vs_p}) with $EM = 10^{50}$\,cm$^{-3}$ yields a
very large filling factor, $f \sim 80$\%. Although not dominant
in terms of total volume emission measure, the cool loop population
possibly covers most of the visible hemisphere.

\section{Summary and conclusions}
\label{sec:concl}

We have performed a detailed analysis of the {\it Chandra}/LETG
spectrum of the dMe flare star AD~Leo, with the aim of deriving a
distribution of the plasma emission measure vs.\ temperature, estimates
of the plasma density at various temperatures by means of spectroscopic
diagnostics, and measurements of the coronal abundances. We have employed these new
results to test our knowledge of the
corona of AD~Leo, based on previous observations with more limited information
content, and to improve our understanding of the thermal
and spatial structuring of the coronal plasma in active red dwarfs with a
``saturated'' X-ray emission level. 

% \begin{enumerate}
% \item
The X-ray emission of AD~Leo during the observation showed
significant low-level variability, typical of such an active flare star.
% Since no large scale flaring event is evident, 
We have analyzed the
X-ray spectrum accumulated during the entire observation,
in order to derive information on the ``average'' properties of the corona
in its characteristic unsteady, but quiescent state.

% \item
The X-ray spectrum of AD~Leo, dominated by emission lines from
\ion{O}{7}--\ion{}{8}, \ion{Ne}{9}--\ion{}{10},
\ion{Si}{12}--\ion{}{14}, and \ion{Fe}{17}--\ion{}{21} ions, allowed us
to reconstruct the emission measure distribution (EMD) in the
temperature range $10^{5.8}$--$10^{7.5}$\,K and to estimate the abundance
ratios of the most common $\alpha$ elements with respect to iron.

% \item
A crucial aspect of our analysis concerns the selection of the
line measurements to be employed for the reconstruction of the EMD.
We discarded the \ion{Ne}{10} Ly$\alpha$ line because it
appears a factor $\sim 2$ less intense than predicted by the EMD, with
respect to the other Ne lines of the Lyman series. This effect could be
due to resonant line absorption or to processes of population of the
atomic levels not yet taken into account properly by the available
plasma emissivity codes (see the case of the \ion{O}{8} 
Ly$\alpha$/Ly$\beta$ ratio presented by Smith et al.\ 2001).

% \item
The EMD peaks at $T \approx 8$\,MK and
can be approximated with a power law with slope $2 \pm 1$, 
in the temperature range 2.5--8\,MK; a secondary maximum occurs at
2\,MK, and a substantial fraction of the emission measure is present also at
temperatures beyond the peak, up to $T \approx 30$\,MK. The shape of the
EMD is in good agreement with that derived from previous analysis of
EUVE spectra: this result strengthens our confidence in the robustness
of a line-based approach for the emission measure analysis and confirms
the ``stability'' of the thermal configuration of the corona of AD~Leo
on time scales of several years. In particular, the bulk of the EMD
appears to be continually present, unrelated to any evident flaring phase.
Flares with medium or large amplitudes likely have significant effects
only on the hot tail of the EMD.

% \item
While the emission measure peak occurs at approximately the same
temperature as in the EMDs of previously studied active binaries or 
single G-K dwarfs and Hertzsprung gap giants,
the shape of the EMD is significantly shallower than that generally 
obtained in these other cases.  With only this single example, 
we cannot establish whether this feature is characteristic of just dMe stars, 
or of stars with medium-low X-ray luminosity ($L_{\rm x} \approx
10^{28}$\,erg s$^{-1}$). At variance with the case of other high-luminosity
active stars, the slope of the EMD is in agreement (within uncertainties)
with that predicted by models of coronal loops shorter than the pressure
scale height, with constant cross-section and uniform heating. 
In particular, the present EMD is
well matched by the static loop model employed by \cite{smf99} to fit the
BeppoSAX spectra of AD~Leo. On the other hand, the slope is also
in agreement with that predicted by the model of a corona entirely
heated by continuously flaring loops, developed by \cite{g+03}.

% \item
In order to test the quasi-static corona model against 
the continuous flaring model, we have considered estimates of the plasma
pressure at various temperatures, obtained by means of density-sensitive
line ratios from He-like triplets and \ion{Fe}{21} lines. The computed
pressures appear to increase with increasing temperature, a result which
we interpret as evidence of different classes of coronal structures
that are not 
in pressure equilibrium.  In particular, while the \ion{O}{7}
triplet yields pressures of 0.5--10\pu at 2\,MK, the
\ion{Fe}{21} lines suggest $p \approx 10^3$--$10^4$\pu.
This result also has relevance to the interpretation of the EMD in
terms of quasi-static loops.  For loops of constant cross-section, 
$EM(T) \propto T^{3/2}$; as we noted earlier, this power law is similar
to that observed.  However, a loop model matching the hot part of
the EMD and containing gas at the high pressure found at these
temperatures also requires the same high pressure at lower temperatures.  
% This situation is ruled out by the much lower pressure measurements 
% based on \ion{O}{7} at 2\,MK. 
The disparity between the different pressure estimates at 2\,MK and 10\,MK
requires that the hot high-pressure loops
yield a negligible contribution to the plasma emission measure at
2\,MK; in turn, this result implies that the EMD of the hot loops should be 
sufficiently steep, possibly steeper than constant cross-section
uniformly-heated models allow.
% loops that are much more isothermal than constant cross-section models allow.  

If we assume that the bulk of the X-ray emission comes from static loops
with high-pressure plasma, the EMD requires very small loop structures 
($L < 10^{-2} R_*$) whose footpoints cover a tiny fraction ($f \ll 1$\%) 
of the visible hemisphere; 
these model loops are significantly shorter than the loops
undergoing flaring events, studied by \cite{fmr00}, which have $L \approx
10^{-1} R_*$. Following this hypothesis leads to
a scenario in which the corona consists of a steady population of
short, high-pressure loops, while the large flares are due to another 
population of longer loops, possibly those suggested by the hot tail of the EMD.
% Alternatively, if we try to interpret the EMD with loops of the same
% size as those undergoing flares, plasma pressures of 10--100\pu are
% implied and surface filling factors of few percent are required.

The small surface filling factor of the main loop population can be
easily reconciled with the lack of any large-amplitude variability of
the X-ray emission: if the loops are uniformly distributed
over the stellar surface, we do not expect any clear rotational
modulation of the coronal emission, and on the other hand, in case of
flares, the small size of these loops ($L \sim 10^8$\,cm) implies very short 
thermodynamic cooling times 
($\tau_{\rm c}$ of the order of 10\,sec according to equation \ref{eq:coolt}),
% \begin{equation}
% \tau_{\rm c} \approx 12 \left( \frac{L}{10^8} \right) 
% \left( \frac{T}{10^7} \right)^{1/2} ~ {\rm sec,}
% \end{equation}
i.e. the X-ray emission from these loops is so rapidly variable that 
no evident flare can be discerned in the light curve.
This argument may also explain why the loops associated with {\em visible} flares 
are preferentially found to be relatively larger.

% \item
Finally, the derived coronal abundances, relative to solar values, tend
to increase with the First Ionization Potential, as in other active
stars. A puzzling result is that the absolute iron abundance in the corona
of AD~Leo derived here, based on the observed line to continuum ratio, 
suggests a nearly solar
metallicity of the plasma, contrary to previous findings from low
resolution X-ray spectra that indicated
$[Fe/H] \sim -0.6$.  Notable results for individual elements include
enhancements relative to Fe of S by a factor of about 7, of Ne by a
factor of 5, and of Si by a factor of 2.5.

% \end{enumerate}

\acknowledgements{AM, SS, GM, and GP acknowledge partial support from
Ministero dell'Universit\'a e della Ricerca Scientifica.
VK and JJD acknowledge support from NASA contracts NAS8-39073 and
NAS8-03060 to the {\em Chandra} X-ray Center; JJD also thanks the NASA
AISRP for providing financial assistance for the development of the
PINTofALE package under NASA grant NAG5-9322. FRH acknowledges partial
support from NASA contracts NAS8-38248 and NAS8-01130.
}

% \newpage
\appendix
\twocolumn
\section{Data analysis and emission line spectroscopy issues}

In this appendix we describe some details of our
spectral analysis of the X-ray emission of
AD~Leo, its
reliability and robustness and
the limitations of our line-based approach.

\subsection{Line identification and fitting}
\label{sec:id}

The line identification list in Table~\ref{tab:lines} is not intended 
to be complete.
We have instead focussed our attention on significant features 
that could be useful in the emission measure analysis.
In the 36--66 \AA\ region~-- where
most of the 3rd order spectrum falls -- several features have escaped 
identification, as have with a few more at long wavelengths 
($\lambda > 80$\,\AA) where the spectrum is more noisy.

The most intense lines are those from the (unresolved) \ion{O}{8} Ly$\alpha$
doublet ($\lambda 18.967,18.972$), the \ion{Ne}{10} Ly$\alpha$ 
doublet +\ion{Fe}{17} blend at 12.12\,\AA, 
the \ion{Ne}{9}+\ion{Fe}{19} blend at 13.45 \AA, the \ion{Fe}{17}
line at 15.02\,\AA\ and the resonance line of the \ion{O}{7} triplet at
20.60\,\AA. 

Among the identified iron lines, 
note in particular the several \ion{Fe}{22}--\ion{Fe}{24} lines
in the region between 10.6\,\AA~and 11.8\,\AA, which also includes
many weak \ion{Fe}{17}--\ion{Fe}{19} lines not
listed in the {\sc chianti} database but present in the Astrophysical
Plasma Emission Database ({\sc aped}).

\begin{figure}[!ht]
   \plotone{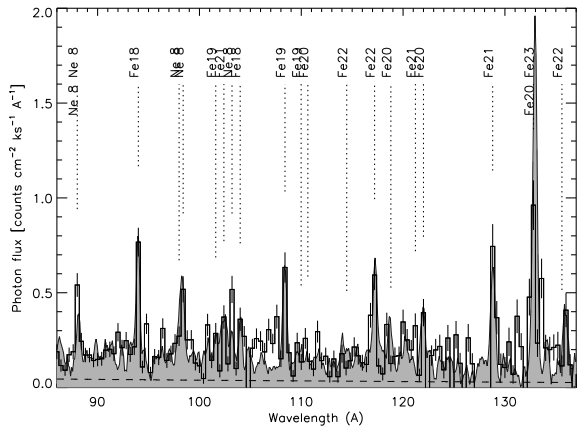}
%  \centering
%  \resizebox{\hsize}{!}{
%  \includegraphics[bb=54 400 558 750]{f10.eps}
%  }
\caption{
Long wavelength region of the AD Leo LETG spectrum (thick line
histogram with error bars), compared with the total EUVE spectrum (thin
line, shaded) obtained as the sum of nine observations. 
The {\it Chandra} spectrum has
been rebinned to match the EUVE spectral resolution (FWHM $\sim 0.4$\,\AA), 
and both spectra were converted to absolute photon flux units
(i.e., divided by the respective instrument effective areas). 
\label{fig:longw}}
\end{figure}

At long wavelengths (80 -- 140\,\AA) we have identified several
\ion{Ne}{8} and \ion{Fe}{18}--\ion{}{23} lines observed in
previous EUVE spectra: Fig. \ref{fig:longw} shows a comparison between
the present {\it Chandra}/LETG spectrum and the exposure-weighted sum of 
the nine EUVE spectra taken between 1993 and 2000 \citep{sm02}. The
{\it Chandra} spectrum has been smoothed with a bin size of 0.4\,\AA\ to match
the FWHM resolution of the short-wavelength (SW) EUVE spectrometer,
and both spectra are in calibrated absolute flux units (having taken
into account the instrument effective area). There is a remarkable
overall agreement between the two data sets, which secures most of the
identifications we have made in the {\it Chandra} spectrum. In particular,
we note the detection of the density-sensitive lines
from \ion{Fe}{20} at $\lambda$ 110.63\,\AA, \ion{Fe}{21} at $\lambda$
121.21\,\AA, and \ion{Fe}{22} at $\lambda$ 114.41\,\AA, which hint
at the presence of high-density plasma 
($N_{\rm e} \ga 5 \times 10^{11}$\,\du) in the corona of AD Leo.

\begin{figure}[!ht]
\plottwo{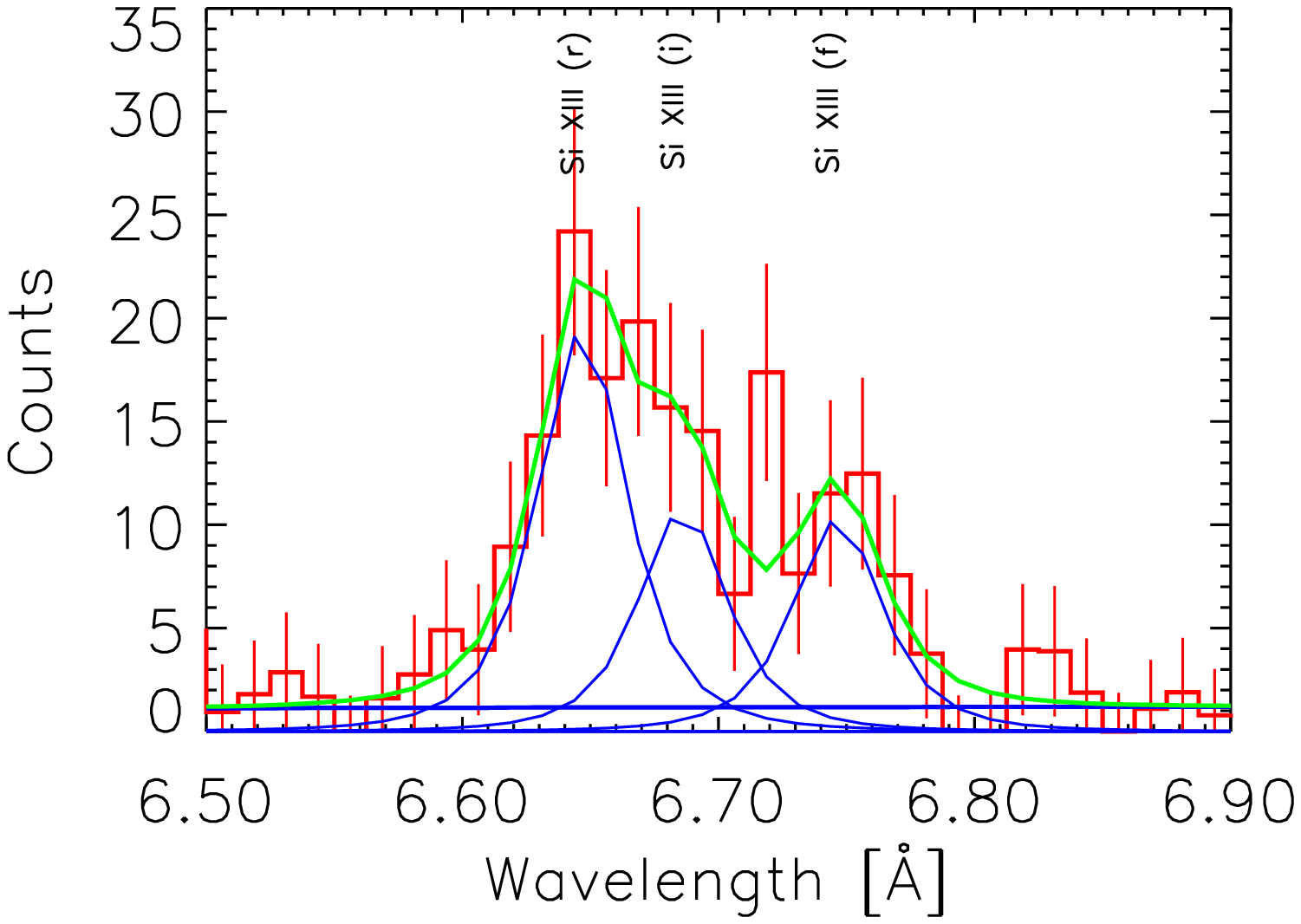}{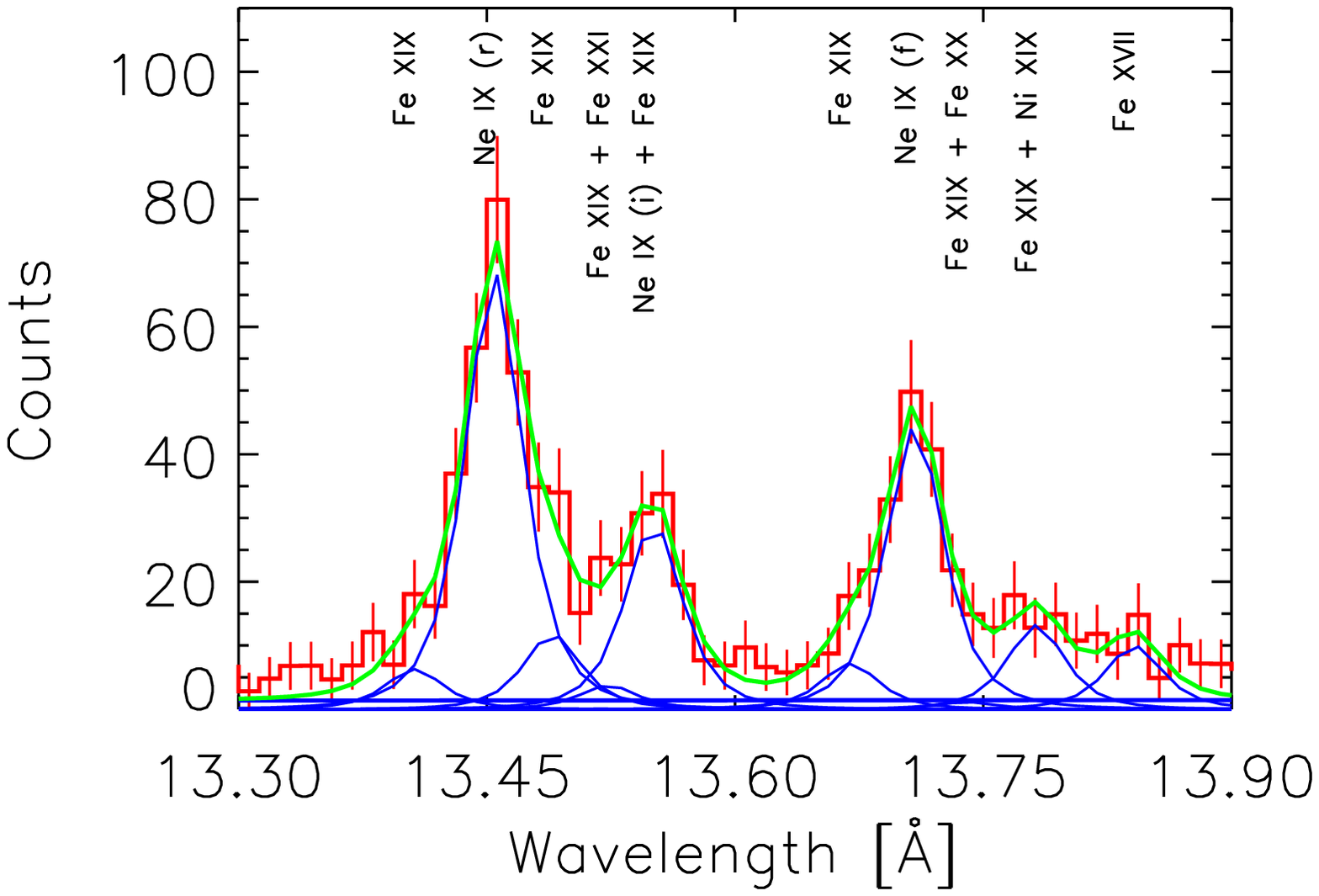}
\caption{
\ion{Si}{13} and \ion{Ne}{9} He-like triplets. The two
panels show the components used to de-blend the line
complexes, and the continuum level (in the electronic edition, the
individual line components and the continuum are colored blue, while the
total model spectrum is green). Note the mixture of
Ne and Fe lines in the wavelength range of the \ion{Ne}{9} triplet.
\label{fig:blends}}
\end{figure}

Central wavelengths and integrated line fluxes have been determined
by line fitting using Moffat profiles, i.e., a modified Lorentzian 
functional form,
\begin{equation}
L_{\beta}(x) = \frac{A}{\left[ 1 + ( (x-x_0)/\sigma )^2 \right]^{\beta} }
\end{equation}
plus a piece-wise constant base level, adjusted to match
the source continuum predicted by the final emission measure
distribution vs. temperature.
The exponent $\beta=1.8$ and $\sigma=0.035$ in the above equation were
determined by fitting the most intense and isolated features in the spectrum
(\ion{Ne}{10}+\ion{Fe}{17} $\lambda$ 12.12 \AA, 
\ion{Fe}{17} $\lambda$ 16.78 \AA, 
\ion{O}{8} $\lambda$ 18.97 \AA, \ion{O}{7} $\lambda$ 21.60 \AA, 
\ion{C}{6} $\lambda$ 33.73 \AA), and then keeping these parameters fixed for 
all the other lines.
Line fitting was performed using multiple components where required by
recognized line blends, and errors at the 68\% confidence level have been also evaluated for each single 
interesting parameter ($\Delta \chi^2 = 1$).
Special attention was paid in fitting the He-like triplets from
the \ion{Si}{13} and \ion{Ne}{9} ions, and the better resolved
triplet of the \ion{O}{7} ion. 
Results of these fits are shown in Fig.~\ref{fig:blends}. 
The \ion{Mg}{11} triplet has been also detected, but it is relatively weaker
and more noisy than the others, and line fluxes determined therefrom have
quite large uncertainties. 

\begin{figure}[!ht]
\plotone{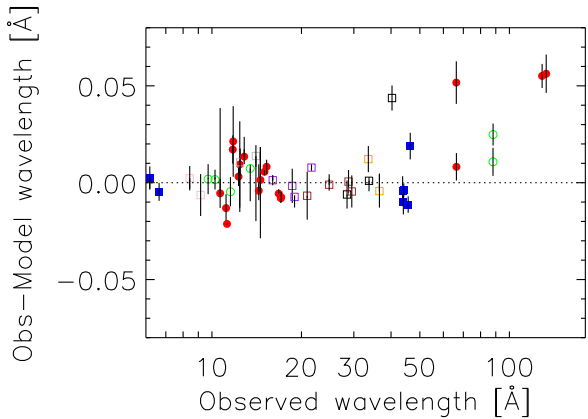}
\caption{
Differences between the measured 
(best-fit) wavelength (with associated 1$\sigma$ error) and the nominal value 
(based on the {\sc chianti} identifications), plotted vs.\ the measured 
wavelength, for the strongest lines in the spectrum
(filled circles:Fe; open circles: Ne; filled squares: Si; open squares: others; in the electronic edition, different elements are colored as in
Fig.~\ref{fig:em}).
% (color code: black C, brown N, purple O, green Ne, blue Si, orange S, red Fe).
The agreement is within 0.03 \AA~
in all cases except a few lines at wavelengths $\lambda > 40$\,\AA,
likely due to residual uncertainties in the wavelength scale.
\label{fig:wvl}}
\end{figure}

Comparison between observed and model wavelengths, i.e., between the fitted 
line center values and the nominal values from the {\sc chianti} 
line identification list, are shown in Fig.~\ref{fig:wvl}: 
the agreement is generally good, within 0.02\,\AA,
except for a positive shift by up to 0.05\,\AA~for few
lines at wavelength $> 40$\,\AA, likely due to residual uncertainties in
the wavelength calibration.

\subsection{Emission measure distribution (EMD) and element abundances}
\label{sec:em}

Emission measure analysis and abundance determination were performed 
in several steps. First, following \citet{p64}, we constructed inverse emissivity curves
for all the detected iron lines,
with the aim to select a subset of them complying with the criterion that
the inverse emissivity curves for lines of the same ion should agree. Single
discrepant lines were discarded, and we have verified that 
in most cases they are affected by problems like heavy blending 
with other known lines, anomalous
profiles, low S/N ratio, or dependence on the plasma density. 
Especially problematic are the \ion{Fe}{18} and \ion{Fe}{19} lines:
among the 14 \ion{Fe}{18} lines with more than 20 integrated counts,
only two ($\lambda\lambda 14.37,14.55$\,\AA) have been included in the
final selection because most of the others have measured fluxes that
differ from those predicted by more than a factor 2 (i.e., the
observed line ratios do not agree -- in most cases -- with those based on
the {\sc chianti} emissivities and the final EMD).
Furthermore, none of the \ion{Fe}{19} lines was selected because they are 
either severely blended or are too weak for reliable line fitting.

We then performed a similar selection for 
the lines of the other elements, one at a time, in the following 
order: Si, Mg, Ne, O, N, C, and finally S and Ni. This sequence is such that 
the line emissivities always have some temperature range in common 
with those of the preceding group, and hence the relative abundance of
each element can be guessed by forcing the overlap of its inverse emissivity 
curves with those of the other elements in that range. 
With the above procedure we
selected about half of the lines in our list, thereby obtaining a first, 
rough estimation of the element abundances with respect to Fe.

The next step was to apply the Markov-Chain Monte Carlo (MCMC) method of
\citet{kd98} to derive consistently the EMD, the
element abundances, and the associated uncertainties. 
This method essentially performs a search in the parameter space
(comprising a set of emission measure values, defined over a pre-selected
temperature grid, and a set of element abundances depending on the
chosen emission lines) while attempting to maximize the probability that the
observed and predicted line fluxes are drawn from the same parent
population.
An important
characteristic of this method is that formal statistical uncertainties
on individual free parameters can be obtained from the distribution of
values in the sampled region of the parameter space near the
final maximum-probability solution. In practice, uncertainties are
available only for a subset of the emission measure values among those
in the EMD, depending
on the availability of observed lines with sensitivity adequate to
each selected temperature.

A critical feature of this method is the choice of the temperature range
where the EMD is reconstructed: it should be large enough to cover the
tails of the emissivity functions of the selected lines with the lowest and 
the highest formation temperature, because significant
contributions to the integrated line flux may come from substantial
amount of plasma located relatively far from the emissivity peak.
On the other hand, since the emission measure values at the extremes 
of the temperature range tend to be
determined by fewer and weaker lines than the values near the peak of
the EMD, the uncertainties on the EMD tails become larger and
larger for increasing width of the explored temperature range.
We have performed two different tests to test the reliability of the
EMD tails: a careful comparison of the model spectrum, based on the
full EMD, with the observed spectrum at relatively long wavelengths 
(20--100\,\AA) can reveal whether or not the emission measure at the 
low-temperature end is overestimated, because in that case several 
low-ionization lines (e.g.\ \ion{Si}{10}, \ion{Si}{11}) are predicted
but not observed. Instead, the EMD high-temperature tail
affects the {\em shape} of the continuum in the critical short-wavelength
range 5--9\,\AA, and comparison with the observed spectrum in this
range has allowed us to test the reconstructed EMD.
Based on the above two tests, we have chosen an optimal temperature
range for the emission measure analysis ($\log T \sim$ 5.8--7.5\,K).
The same tests also suggest that the uncertainties on the emission
measure near the extremes of the temperature range are likely larger
than the formal values determined in each single run of the MCMC
procedure.

A second delicate aspect of the analysis is the line selection, and in
particular the possible bias introduced by the strongest resonance
lines, namely the Ly$\alpha$ lines from hydrogenic ions and the
iron lines which form near the peak of the EMD (\ion{Fe}{17} in the
case of AD~Leo). As a matter of fact, the accuracy of the atomic
databases and of the model EMD we derive under the usual (wide) set of
assumptions (optically-thin emission from a plasma in ionization
equilibrium with uniform abundances) is such that we are currently able
to predict most of the observed emission lines, but not better than to
within a factor 2. This systematic uncertainty is in many cases much
larger than the statistical uncertainty on the measurements,
especially for the strong lines mentioned above, with the consequence
that the emission measure analysis can be adversely influenced by
the attempt to match the observed flux of these lines. The case of the
\ion{Ne}{10} Ly$\alpha$ line + \ion{Fe}{17} is revealing: this is the
strongest spectral feature among those dominated by Ne lines,
and -- when it is included in the selected line list -- the MCMC procedure
yields an EMD and Ne/Fe abundance ratio that predict the observed
flux of this line almost perfectly. With this choice, however, 
all other
Ne lines are systematically over-predicted, with the isolated
\ion{Ne}{10} Ly$\beta$ line at 10.24\,\AA\ over-predicted by more than
a factor 2. With the Ly$\alpha$ line instead excluded, the new EMD and Ne/Fe
value provide a fair prediction of all the Ne lines (Fig. \ref{fig:flx}),
but the observed flux of the Ly$\alpha$ line is lower than
predicted by a factor $\la 2$. Possible reasons for this behavior
were discussed in Sect.\,\ref{sec:a+r}.

Finally, we discuss the issue of missing lines in the {\sc chianti}
database. As discussed by \citet{amp03} and by \cite{smp+04},
the {\sc chianti} database is currently less complete than the {\sc aped}
database, especially in the wavelength region 10--12\,\AA\ where
hundreds of weak lines from \ion{Fe}{17}--\ion{}{19} fall. This
spectral region also includes the most relevant L-shell lines 
from \ion{Fe}{22}--\ion{}{24}, which are crucial to constrain the
high-temperature tail of the EMD. Most of these lines are blended with
iron lines of lower ionization stages, whose contributions need to
be taken into account. Due to the incompleteness of the {\sc chianti}
database, we have adopted the following approach: the EMD was initially
reconstructed including only lines in the {\sc chianti} database; then, we
synthesized the predicted model spectrum using the {\sc aped} V1.3 
database, and evaluated any possible (previously missed) contribution to 
the relevant \ion{Fe}{22}--\ion{}{24} lines. The measured line fluxes of
these latter lines were then multiplied by a correction factor ($< 1$, essentially a
blending fraction), and the
emission measure analysis was iterated until convergence was achieved.
The affected lines with their estimated blending fraction, based on
{\sc aped}, are as follows: \ion{Fe}{24} $\lambda 10.66$ (60\%
contamination), $\lambda 11.17$ (70\%),  $\lambda 11.26$ (90\%). A
similar correction was also applied to the \ion{Ne}{9} line 
at $11.55$\,\AA\ (30\% contamination).
We stress that, if the above correction is not applied, the
overestimated strength of the \ion{Fe}{22}--\ion{}{24} lines
yields an EMD with a more pronounced high-temperature tail.
Most of the high-ionization Fe lines remain under-predicted by
the EMD model, however, because the EMD at $T \gtrsim 10^7$\,K cannot grow
arbitrarily, due to the
constraints provided by other lines from \ion{Fe}{21}, \ion{Fe}{23}, \ion{Mg}{12},
and \ion{Si}{14} ions (Fig.~\ref{fig:em}).

\subsection{Metallicity}
\label{sec:metal}

We have identified two wavelength regions where a comparison between
the observed spectrum and the predicted continuum emission level is
especially useful for determining the plasma metallicity: 
the range 6--9 \AA, where the cutoff of the
bremsstrahlung emission from the hottest plasma ($T \sim 2 \times
10^7$\,K) occurs, and the region long-ward of the C K-edge 
($\lambda \gtrsim 44$\,\AA), where the jump in the instrument effective area 
makes the continuum visible again.
A close inspection of Fig.~\ref{fig:z} reveals that, in both regions,
a reasonably good agreement can be reached only for $Fe/H$ ratios
$\ge 0.8$ the solar value. The relatively low continuum level, due to
this metallicity value, makes it difficult to constrain the iron
abundance: in fact, an increase of $Fe/H$ by a factor 2 
(i.e., from $Fe/H = 0.8$ to 1.6 times solar) implies a small change 
in the predicted continuum level, 
comparable to the uncertainty in the observed count rate 
due to Poisson statistics.
For this reason we have conservatively deduced for AD Leo a
metallicity comprised in the range 0.8 -- 1.6 times the solar photospheric
value ($\log A_{\rm Fe} = 7.51 \pm 0.05$, where $\log A_{\rm H} = 12$).
Such a metallicity determination implies that a large fraction of the 
apparent continuum between 5 and 15\,\AA~is actually due to emission lines. 

\subsection{Density and temperature diagnostics}
\label{sec:dens}

The \ion{O}{7} He-like triplet is the only well-resolved triplet in
{\it Chandra}/LETG spectra. Nonetheless, two issues are relevant for a
correct estimation of the plasma (effective) density and temperature:
the placement of the continuum level, which affects
the measurement of the line fluxes, as demonstrated by \citet{b+02},
and the proper treatment of transitions from high-$n$ states
\citep{pmd01,sbl+01}. The continuum we have used is that
predicted by the EMD, and hence it should not suffer of misplacement
problems due to unresolved weak lines in the relevant wavelength
range. In order to check for possible deficiencies in the {\sc chianti}
database, we compared our estimated density and temperature with those
predicted by the model of \citet{pmd01} for the same measured \ion{O}{7}
line ratios and found complete agreement between them. 
Our result is also consistent with that obtained by
\citet{nsb+02}, based on the latter model.

More problematic is the proper measurement of the line fluxes in the
\ion{Ne}{9} triplet, because of the heavy blending with \ion{Fe}{19}
lines, and to a lesser extent also \ion{Fe}{20}--\ion{}{21} lines.
\citet{nbdh03} have recently performed a detailed analysis of the 
\ion{Ne}{9} triplet spectral region in
{\it Chandra} and XMM-Newton spectra of Capella, and concluded that the
Astrophysical Plasma Emission Code (APEC) models are sufficiently accurate and complete that all
significant observed lines in this region can be reasonably identified
in the {\it Chandra}/HEG; they also argue that spectra obtained by LETGS are 
inadequate to derive reliable results independently. 
In order to measure the line
fluxes in our AD~Leo spectrum we have performed a multi-component line
fitting by taking into account most of the spectral features listed by
\citet{nbdh03}. Moreover, for the resonance, intercombination, and
forbidden \ion{Ne}{9} lines we have evaluated the (density-dependent)
contamination due to unresolved iron lines, as predicted by the model
spectrum and both the {\sc chianti} and {\sc aped} databases. In
Table~\ref{tab:ne9} we report the blending fraction in the low-density
limit for both atomic databases, and at the estimated plasma density
in the corona of AD~Leo for {\sc chianti} only. While the contamination of
the resonance line is the same in all cases, the blending fraction for
the intercombination line decreases for increasing density (because
only the Ne contribution is density-dependent), and the opposite
occurs for the forbidden line, as expected. Rather surprising is
the finding that in the low-density limit, the {\sc chianti} database predicts
a larger blending fraction than {\sc aped}.

\begin{figure}[!ht]
\plotone{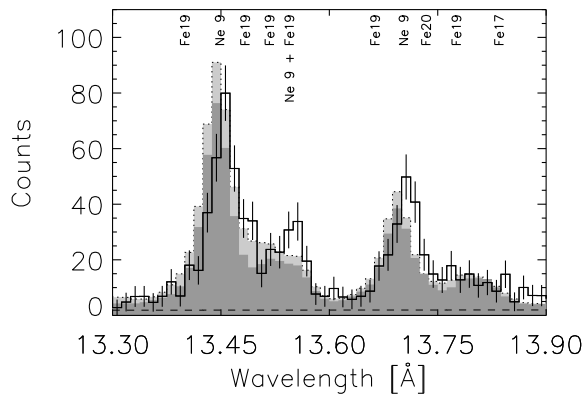}
\caption{
The \ion{Ne}{9} triplet spectral region with model predictions in the
low-density limit based on the reconstructed emission measure
distribution and the {\sc chianti} (dark shade) or {\sc aped} (dot-bounded
light shade) line emissivities. 
The excess emission observed at $\lambda \sim 13.5$\,\AA\
is suggestive of a high-density effect.
\label{fig:ne}}
\end{figure}

In order to compute the relevant density- and temperature-sensitive
ratios we subsequently corrected the
measured line fluxes, in a self-consistent way, for the blending fraction computed according to
{\sc chianti}. Both the temperature and density determinations
(Table~\ref{tab:tri}) are in agreement with the alternative estimates
based on the \citet{pmd01} model. However, at variance with
\citet{nsb+02}, our analysis suggests a plasma density significantly
above the low-density limit: in fact, Fig.~\ref{fig:ne} shows excess
emission at the location of the intercombination line with respect to
the low-density predictions based either on {\sc chianti} or {\sc aped}. On the
other hand, both model spectra show locations of the emission peaks 
(corresponding to the resonance and forbidden lines) that are shifted
by one bin (0.0125\,\AA) with respect to the data, 
which we interpret as evidence of uncertainties in the
atomic databases related to the wavelength of the emission lines
or to the relative emissivities, or both.

For the \ion{Si}{13} triplet we followed a similar procedure. 
In this case, a minor contamination (blending
fractions $< 4$\%) to the intercombination and forbidden lines 
is due mainly to \ion{Si}{12} satellite lines. The {\sc chianti} density and
temperature estimates are again compatible, within uncertainties, with
those based on the \citet{pmd01} model. However, some difficulty with the
atomic models arises from the finding that the \ion{Ne}{9} and
\ion{Si}{13} temperature-sensitive
ratios, $R_{\rm T}$, are quite high, yielding estimates of the plasma
temperature that are significantly lower than the average values 
weighted by the line emissivity function and by the EMD. This puzzling
result was recently noted in other contexts \citep{amp03,nbdh03}, but 
a clear explanation is not yet available.

Finally, we looked for density-sensitive \ion{Fe}{20}--\ion{}{22}
lines in the AD~Leo spectrum. These lines provide independent
estimates of the plasma density at temperatures ($T \sim 10^7$\,K) 
similar to those at which the \ion{Si}{13} triplet forms, and they have
been commonly employed in studies of EUVE spectra \citep{sm02}.
In the AD~Leo spectrum we tentatively identified the lines
\ion{Fe}{20} $\lambda 110.63$, \ion{Fe}{21} $\lambda 121.20$, and
\ion{Fe}{22} $\lambda 114.41$, which become clearly visible only
in high-density plasmas ($N_{\rm e} \ga 10^{11}$--$10^{12}$\du).
These lines have only a few tens of integrated counts, and the relevant
spectroscopic diagnostics (provided by the line ratios 
110.63/121.84, 121.2/128.75, 114.41/117.18) yield density estimates
affected by large error bars. The most precise result is that
derived from the \ion{Fe}{21} line ratio and reported in
Table~\ref{tab:tri}, suggesting a plasma density $N_{\rm e} \sim
10^{12}$\du.

% \clearpage

\begin{deluxetable}{cccc}
\tablecaption{Blending fractions\tablenotemark{a} ~for the \ion{Ne}{9} triplet
\label{tab:ne9}}
\tablecolumns{4}
\tablewidth{0pt}
\tablehead{
 & \multicolumn{1}{c}{APED} & 
\multicolumn{2}{c}{CHIANTI} \\
\colhead{Line} & 
\colhead{Low-density} & 
\colhead{Low-density} &
\colhead{$5 \times 10^{11}$\du\tablenotemark{b}} \\
}
\startdata
r  & \phn6\% & \phn6\% & \phn6\% \\
i  & 22\% & 31\% & 22\% \\
f  & \phn2\% & \phn3\% & \phn4\% \\
\enddata
\tablenotetext{a}{Fraction of flux from contaminating Fe lines
in a range of about 0.02\,\AA\ centered on the Ne line.}
\tablenotetext{b}{Estimated electron density from corrected f/i ratio.}
\end{deluxetable}

% \clearpage

% \newpage

\end{document}